\documentclass[3p,sort&compress]{elsarticle}

\usepackage{lineno,hyperref}
\modulolinenumbers[5]

\journal{Journal of Non-Newtonian Fluid Mechanics}

\usepackage[utf8]{inputenc}
\usepackage{xcolor}
\usepackage{graphicx} 
\usepackage{float}

\usepackage{color}
\usepackage{siunitx}

\usepackage{amsmath}
\usepackage{amssymb}

\usepackage{algorithm}
\usepackage[noend]{algpseudocode}

\usepackage{upgreek}

\makeatletter
\def\BState{\State\hskip-\ALG@thistlm}
\makeatother

\begin{document}

\begin{frontmatter}

\title{Computational Rheometry of Yielding and Viscoplastic Flow in Vane-and-Cup Rheometer Fixtures}

\author[STRATH]{Emad Chaparian\fnref{eqau}\corref{mycorrespondingauthor}}
\cortext[mycorrespondingauthor]{Corresponding author: emad.chaparian@strath.ac.uk}
\fntext[eqau]{These authors contributed equally to this work}
\author[MIT]{Crystal E. Owens\fnref{eqau}}
\author[MIT]{Gareth H. McKinley FRS}

\date{\today}

\address[STRATH]{James Weir Fluid Laboratory, Department of Mechanical \& Aerospace Engineering, University of Strathclyde, Glasgow, United Kingdom}
\address[MIT]{Hatospoulos Microfluids Laboratory, Department of Mechanical Engineering, Massachusetts Institute of Technology; Massachusetts, USA}

\begin{abstract} 
A planar two-dimensional computational analysis is presented to qualify traditional and fractal vane-in-cup geometries for accurate rheometry of simple viscoplastic fluids with and without slip. Numerical simulations based on an adaptive augmented Lagrangian scheme are used to study the two-dimensional flow field of yield-stress fluids within and around vane tools with $N=3$ to $24$ arms for a wide range of Bingham numbers, $\mathcal{B}$ (i.e.~the ratio of the yield stress over the characteristic viscous stress). This allows for accurate calculations of the velocity and stress fields around vanes with various geometries, as well as direct comparison to experimental observations of the  output torque measured by a rheometer, revealing sources of variation and error. We describe the impact of the vane structure on the fluid velocity field, from few-arm cruciform vanes ($N\leq6$) that significantly perturb the flow away from ideal azimuthal kinematics, to many-arm fractal vanes ($N\geq12$) in which the internal structural features are successfully ``cloaked" by a yield surface. This results in the shearing of an almost-circular ring of viscoplastic fluid that is indistinguishable from the annular ring of fluid deformed around a slip-free rotating cylindrical bob and leads to more accurate rheometric measurements of the material flow curve. Moreover, in direct comparison with data from previous literature, we show that slip conditions on the vane surface do not impact the velocity field or measured overall torque $\mathcal{T}$, whereas slip conditions on the smooth outer wall have significant impact on data, even when using a vane geometry. Finally, we describe the impact of vane topography and Bingham number, $\mathcal{B}$, on the measured torque and rheometric accuracy of vane-in-cup geometries for ``simple" (inelastic) yield-stress fluids described by either the Bingham plastic or Herschel-Bulkley constitutive model. 

\end{abstract}


\end{frontmatter}


\section{Introduction}
A diverse range of important complex fluid materials exhibit yield stress behavior, including concrete and mortar, foams, foods, cosmetics, drilling muds such as Bentonite solutions, lava, biological gels, 3D and screen printing inks among others. While the presence of the yield stress typically conveys beneficial properties to the fluid such as mechanical and mixture stability or functionality, it simultaneously complicates rheological measurements. The challenging sub-class of complex fluids referred to collectively as `yield-stress fluids' exhibits a wide spectrum of rheological phenomena from linear viscoelasticity (below yield) to nonlinear effects such as shear thinning;  time-dependent behavior such as hysteresis, aging, rejuvenation, and thixotropy \cite{Larson2019, Barnes1999}; as well as more complex elastoplastic effects close to yielding including nonlinear elastic deformation, viscosity bifurcation, delayed yielding, and ``avalanching" flow \cite{Coussot2002_bifurcation, Coussot2002b_avalanche}. Fluid-particle microstructural interactions can also result in shear-induced particle migration \cite{Chow1994_migration}, shear banding, and shear-enhanced settling phenomena. During rheometric measurements, these complex rheological properties must be decoupled from common errors incurred in measurements that are equally numerous. Such errant effects include sample slip on smooth tool surfaces; long-term sensitivity to deformation history; sample loading/tool insertion-induced rejuvenation; trapped bubbles; and unwanted aging, as well as elastic instabilities such as edge fracture \cite{Nguyen1987, Barnes2001, Larson2019, Meeten1992}. These errors are also influenced by the protocol and rheometer tool used to make the measurement. To address such issues the standard cruciform vane rotor tool  was first introduced to rheologists in the 1980s by civil engineers studying the yield stress behavior of bauxite residue (``red mud") \cite{Nguyen1983, Keentok1982}. Due to an array of practical advantages, primarily related to slip prevention and easy insertion into thick fluids, it has subsequently become a well-established tool for measuring the rheological properties of yield-stress fluids, as it provides a good compromise between robust measurements and mitigation of systematic error for these kinds of fluids. 

\begin{figure}[H]
    \centering
    \includegraphics[width=0.8\textwidth]{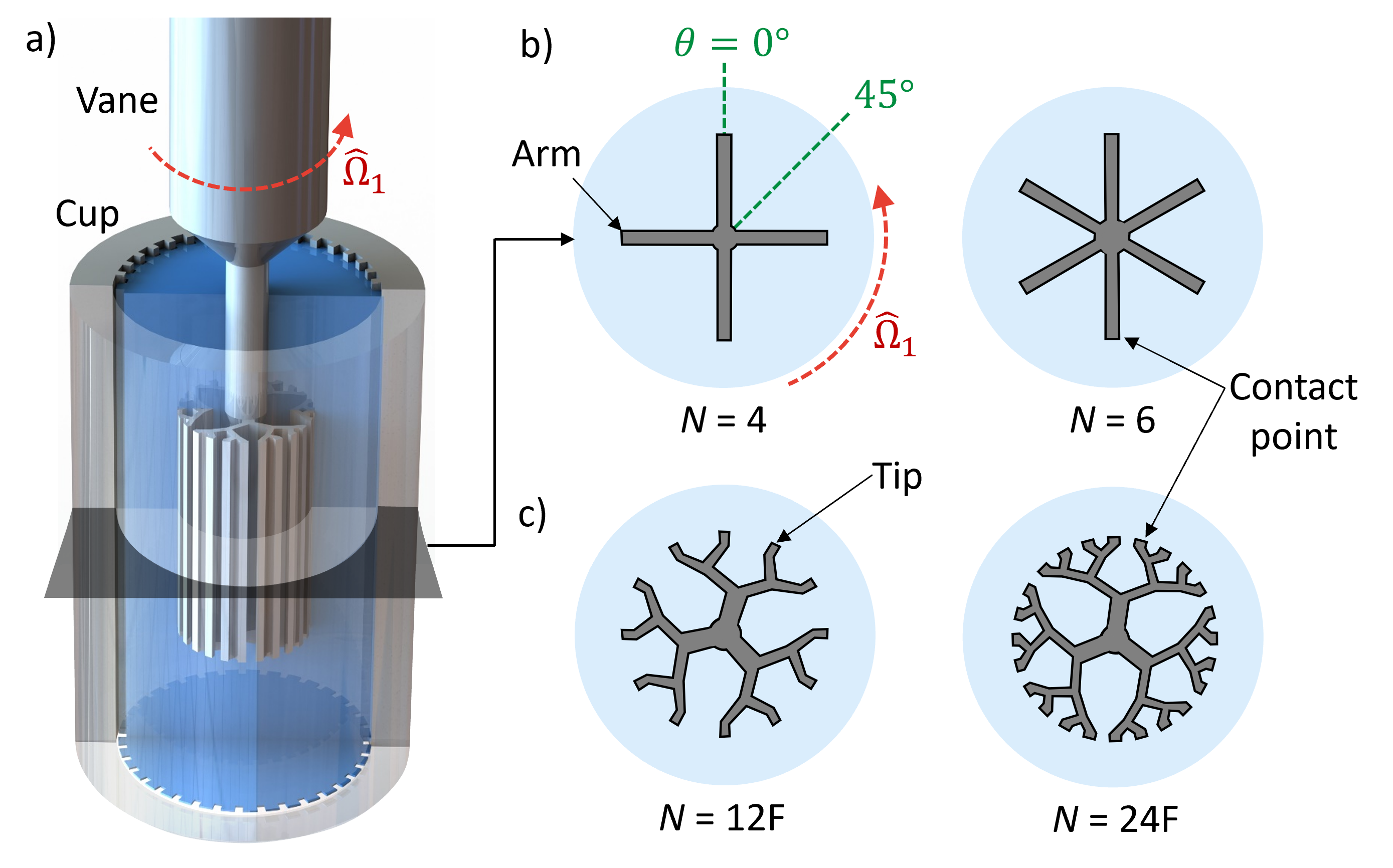}
    \caption{(a) Schematic of a vane-in-cup geometry with a 24-arm fractal vane rotating inside a slip-free cup, which may be implemented by a macroscale grooved texture, as depicted, or by micro-texture or chemical treatment. (b) Cross-sectional images show classical cruciform vanes with $N=4$ and $6$ straight arms, as well as (c) fractal vanes with designs adapted  from Owens et al.~\cite{Owens2020} denoted $N=12F$ and $N = 24F$ with 12 and 24 contact regions respectively.}
    \label{fig1}
\end{figure}

The vane rotor consists of a series of vertical blades fanning out from a central shaft, most commonly having 4 to 8 blades (Fig.~\ref{fig1}(b)) \cite{Barnes2001}. Similar to other axisymmetric geometries, the vane rotates concentrically inside a cup holding the sample fluid (Fig.~\ref{fig1}(a)). In a standard Taylor-Couette geometry, however, viscoplastic samples often slide or slip relative to the smooth tool and cup surfaces, instead of adhering to the surface and satisfying the commonly-assumed ``no-slip” boundary condition \cite{Nguyen1983, Barnes2001}. One key advantage of the vane rotor is that sample fluid is trapped within the rotating vane arms and is forced to rotate as a solid, ideally cylindrical, plug \cite{Barnes2001}. Importantly, if the material is viscoplastic and a yield surface develops; the resulting sample-on-sample shearing surface corresponds to homogeneous yielding and avoids wall slip. Meanwhile, the cup geometry mitigates the severity of edge fracture seen in parallel plate and cone-and-plate geometries by configuring the yielding surface orthogonal to the free surface. Finally, the cross-sectional area of the vane tool (or the \textit{occluded area fraction} shown by the grey regions in Fig.~\ref{fig1}) is sufficiently sparse that it reduces the extent of insertion-induced deformation of   history-sensitive materials, especially when compared to the strong shear experienced when loading fluid samples into the narrow gaps that characterize most other geometries for rheological measurements. For the vanes shown in Fig.~\ref{fig1}, the occluded area fractions are 0.15, 0.20, 0.21, and 0.29 for the 4, 6, 12, and 24-arm vanes, respectively.

Given the similarities of the ideal flow kinematics around a vane to the one-dimensional azimuthal flow within a Taylor-Couette geometry, it is common to use a so-called ``Couette analogy" to derive the equations employed to evaluate material properties (shear stress, shear rate) as a function of the raw values (torque, rotation rate) measured by the rheometer with a vane tool attached  \cite{Medina-Banuelos2019,Nguyen1987}. Using this simplification, the rotating vane is treated as a cylinder with an effective radius, $\hat{R}_{\text{\it eff}}$, which is calibrated by testing the vane with Newtonian calibration fluids. Generally it is found that $\hat{R}_{\text{\it eff}}<\hat{R}_{vane}$, suggesting that there is a small amount of additional sample deformation within the region swept out by the vane arms \cite{Medina-Banuelos2019, Barnes2001}. Throughout this paper we use a `caret' $\hat{(\cdot)}$ to indicate a dimensional quantity. The torque-to-stress conversion factor $S_{couette}$ that is typically used \cite{Nguyen1987, Macosko1994_TCintro, Medina-Banuelos2019} is
\begin{equation}
S_{couette} = \frac{\hat{\tau}}{\mathcal{\hat{T}}} = \frac{1}{2\pi \hat{R}_{\text{\it eff}}^2 \hat{L}(1+\frac{2\hat{R}_{\text{\it eff}}}{3\hat{L}})}
\end{equation}
where $\hat{\tau}$ is the wall shear stress at a radius $\hat{r}=\hat{R}_{\text{\it eff}}$, $\hat{\mathcal{T}}$ is the measured torque, and $\hat{L}$ is the length of the vane. The term $2\hat{R}_{\text{\it eff}}/3\hat{L}$  accounts for ``end effects” due to additional shear stress contributions arising from the two end faces of the finite-length vane \cite{Macosko1994_endeffects}. To diminish this end effect, a typical rheometric vane tool is relatively slender, i.e.~$\hat{L}/\hat{R}_{\text{\it eff}}\geq4$. 

However, contrary to common assumptions, the flow field around a real vane often deviates from the ideal axisymmetric flow profile (i.e.~the assumption of a purely azimuthal velocity field, $v_{\theta}(r)$) for both Newtonian and yield-stress fluids. This deviation has been shown clearly in simulations \cite{Barnes2007} as well as in rheometric experiments using vanes that also incorporate \textit{in situ} measurements of local deformation; examples include particle image velocimetry (PIV) investigations on Carbopol \cite{Medina-Banuelos2019}, as well as magnetic resonance imaging (MRI) studies on emulsions and particle suspensions \cite{Ovarlez2011}. These studies with a range of yield-stress fluids confirmed that the material enclosed within the vane arms experiences some deformation,  instead of rotating ideally as an unyielded plug within the vane. This non-viscometric flow becomes most important at low shear rates, critically disrupting measurements of material yield stress. In addition, Medina-Ba{\~{n}}uelos et al.~showed clearly that for at least one viscoplastic fluid, $\hat{R}_{\text{\it eff}}$ varies with changes in the rotation rate, again breaking effective application of the Couette analogy \cite{Medina-Banuelos2019}.

In order to address some of these stated challenges with non-circularity and empirical calibration, Owens et al.~recently introduced new vane designs with a fractal cross section optimized to create a more axisymmetric shear stress and shear rate field (Fig.~\ref{fig1}(c)) \cite{Owens2020}. They also proposed a torque-to-stress conversion factor that is claimed to not require calibration with a Newtonian fluid or use an experimentally-derived value of $\hat{R}_{\text{\it eff}}$. Instead, the equation uses the physical vane radius $\hat{R}_1$ and incorporates the non-cylindrical shape of the vane by accounting for the number of contact points, $N$, that the vane has with its circumscribed circle. For example, the most common cruciform four-arm vane would have $N=4$. This shape effect is based on an analytical solution for stress imposed on a rotating vane with an arbitrary number of arms \cite{Atkinson1992a}, coupled with experiments with various vane structures to help accurately assess the end effect \cite{Sherwood1991a}. This direct incorporation of the vane shape circumvents the use of the empirically-based Couette analogy. This $N$-dependent torque-to-stress conversion factor is given by
\begin{equation}\label{eq:converttorque}
S_{N} = \frac{\hat{\tau}}{\hat{\mathcal{T}}} = \frac{1}{2\pi \hat{R}_1^2L(1-\frac{1.113}{N})+\frac{\hat{R}_1}{4\hat{L}}(2.75-\frac{3}{\sqrt{N}})}.
\end{equation}
In addition, the conversion of rotation rate to shear rate at the vane perimeter for yield stress fluids in partially yielded flow is given by \cite{Nguyen1987, Owens2020}
\begin{equation}\label{eq:SR}
\hat{\dot{\gamma}}(\hat{R}_1) = \frac{2\hat{\Omega}}{d\log \hat{\mathcal{T}}/d\log \hat{\Omega}}.
\end{equation}
Using this fractal geometry and equation set, Owens et al.~\cite{Owens2020} measured the flow curve of a simple yield-stress fluid, Carbopol, with measurement errors reduced from $>15\%$ for a 4-arm vane to less than $5\%$ by using either the shape-dependent torque-to-stress conversion factor with a 4-arm vane, the novel fractal tool having $N=24$, or both. These results indicate that the accuracy of rheometry on simple yield-stress fluids can effectively be improved both by greater understanding of the flow field used in calculations, and by using tools specifically designed to avoid errors. However, these prior results are based solely on bulk material measurements from a rheometer. A deeper understanding of the local kinematics with spatial resolution near the arms of the vane is needed to confirm the consequences of various design choices, as well as to elucidate, in a general way, the underlying interactions between vane design, viscoplastic fluid properties, and accurate and reproducible measurements.

To address this gap in understanding, we use computational methods to investigate yielding and plastic flow of simple yield-stress fluids in the vane-in-cup configuration. Because we seek to describe a wide range of flow conditions, ranging from the onset of yielding to the fully-developed viscous flow limit, an augmented Lagrangian/adaptive mesh approach \cite{roquet2003adaptive} is utilized to ensure we accurately capture the location and configuration of the yield surface. As shown by Frigaard and Nouar \cite{frigaard2005usage}, simulating the rheological response of an unregularized viscoplastic constitutive model is vital to avoid tracking non-physical solutions when approaching the yield limit, and this is made feasible by using the augmented Lagrangian/adaptive mesh approach. We also investigate the role of `apparent' slip of the viscoplastic fluid over smooth solid surfaces by proposing a new approach that couples the classical augmented Lagrangian method with a slip law \cite{chaparian2020sliding,roquet2008adaptive}.

The outline of the paper is as follows. In Section \ref{sec:ProblemStatement}, we review the mathematical equations and parameters used in the simulations, discuss the boundary conditions and algorithms, and clarify the meshing procedure. In Section III.1, we investigate the effect of the Bingham number on the flow field surrounding a 4-arm vane. We investigate the effect of vane shape on the velocity profiles and yield surfaces for a series of classical vanes having $N = 4, 6$ and $12$ arms, and we investigate the homogeneity of the shear rate within the sheared fluid using the ratio of the extensional and shear components of the velocity field within the fluid domain. In Section III.2, we estimate the contribution of end effects from the upper and lower shearing surfaces of the vane on the torque measured during experiments, and use this relationship to directly compare our two-dimensional (2D) simulations with published experimental results (including PIV and rheometry data \cite{Medina-Banuelos2019}) for a 6-arm vane, using various slip conditions within the simulation to understand their impact on the experimental data. This particular data set was selected for comparison because it employs a simple (non-thixotropic, non-hysteretic) yield-stress fluid, provides spatial velocimetry data (obtained using PIV) alongside standard rheometric data, and includes a thorough analysis of local velocity profiles as well as slip behaviors with carefully measured slip laws, providing a data-rich source for direct comparison to the spatially-resolved simulations presented in the current article. 

In Section IV.1, we compare the performance of slip-free $4$-arm vanes, $24$-arm fractal vanes, and cylinders in terms of the regularity/smoothness and position of the outer yield surface, which show a progressive transition from periodically-modulated to circular yield surfaces with changes in the vane geometry. In Section IV.2, we compare the performance of several vanes for determining accurate flow curves of (nondimensional) wall shear stress versus Bingham number for both a simple Bingham plastic and a more realistic Herschel-Bulkley fluid. A direct comparison to experimental data, appropriately incorporating end effects, shows good agreement between the $24$-arm fractal vane and an ideal slip-free Couette rotor. 

Finally in Section V, we discuss some additional practical experimental considerations regarding tool choice beyond the results of the simulation, and we summarize the most appropriate choice of conversion equations for a given tool using additional measured flow curves from \cite{Medina-Banuelos2019} as a reference case. 


\section{Equations, scalings, and numerical simulations}\label{sec:ProblemStatement}

\begin{figure}
\centerline{\includegraphics[width=0.3\linewidth]{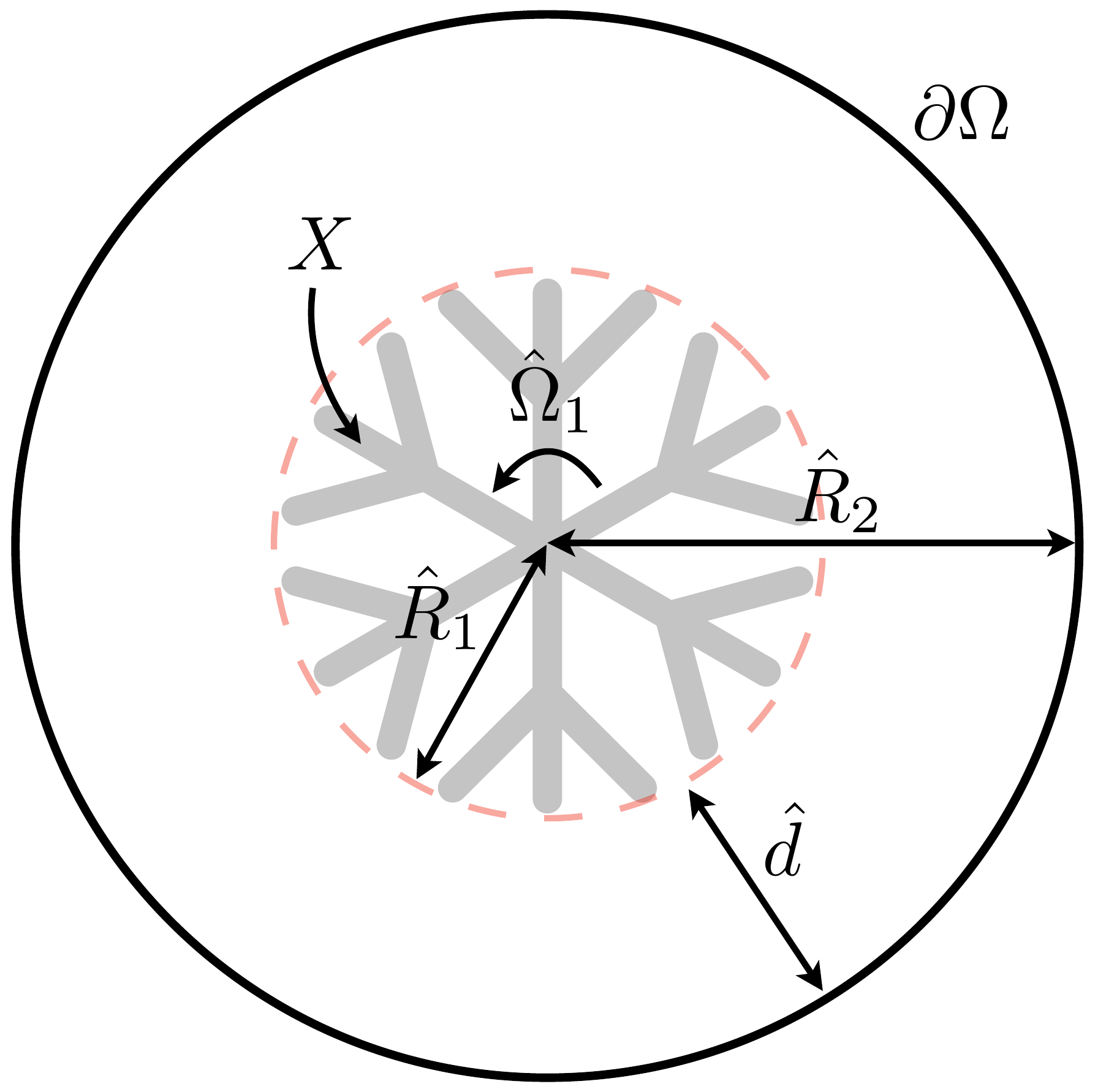}}
\caption{2D schematic of the vane-in-cup geometry. The vane rotates at a rotation rate $\hat{\Omega}_1$. The nominal `gap' width is $\hat{d} = \hat{R}_2 - \hat{R}_1$.}
\label{fig:Schematic}
\end{figure}

We consider steady inertialess flows of a Herschel-Bulkley fluid in a vane-in-cup geometry; see Fig.~\ref{fig:Schematic}. A vane ($X$) of radius $\hat{R}_1$ sits in a cup ($\Omega$) of radius $\hat{R}_2$ filled with the sample fluid. The boundary of the cup and the surface of the vane are denoted by $\partial \Omega$ and $\partial X$, respectively. Hence, the  inertialess Cauchy momentum equation can be written in the form,

\begin{equation}\label{govern}
0 = - \boldsymbol{\nabla} \hat{p} + \boldsymbol{\nabla} \boldsymbol{\cdot} \hat{\boldsymbol{\tau}}~~\text{in}~ \Omega \setminus\bar{X},
\end{equation}
and the constitutive equation is,
\begin{equation}\label{const}
  \left\{
    \begin{array}{ll}
      \hat{\boldsymbol{\tau}} = \left( \hat{K} \Vert \hat{\dot{\boldsymbol{\gamma}}} \Vert^{n-1} + \displaystyle{\frac{\hat{\tau}_y}{\Vert \hat{\dot{\boldsymbol{\gamma}}} \Vert}} \right) \hat{\dot{\boldsymbol{\gamma}}} & \mbox{iff}\quad \Vert \hat{\boldsymbol{\tau}} \Vert > \hat{\tau}_y, \\[2pt]
      \hat{\dot{\boldsymbol{\gamma}}} = 0 & \mbox{iff}\quad \Vert \hat{\boldsymbol{\tau}} \Vert \leqslant \hat{\tau}_y,
  \end{array} \right.
\end{equation}
where $\hat{p}$ is the pressure, $\hat{\boldsymbol{\tau}}$ is the deviatoric stress tensor, $\hat{\dot{\boldsymbol{\gamma}}}$ is the rate-of-strain tensor, $\hat{K}$ is the fluid consistency, $n$ is the power-law index, and $\hat{\tau}_y$ is the yield stress. The maximum radius of the vane is designated by $\hat{R}_{\text{1}}$ (the radius of the tool from the axis of rotation to the tip of the vane), the cup radius by $\hat{R}_\text{2}$, and $\hat{d}=\hat{R}_2 - \hat{R}_1$ is the `nominal gap' width'. Please note that $\Vert \cdot \Vert$ (in Eq.~\ref{const} and throughout the paper) indicates the norm associated with the tensor inner product $\boldsymbol{c} \boldsymbol{:} \boldsymbol{d} = 1/2 \sum_{ij} c_{ij} d_{ij}$, e.g.~$\Vert \hat{\boldsymbol{\tau}} \Vert = \left( \hat{\boldsymbol{\tau}} \boldsymbol{:} \hat{\boldsymbol{\tau}} \right)^{1/2}$.

\subsection{Numerical details}

For viscoplastic flow problems, due to the non-smooth nature of the equations at the yielding point, classical numerical methods (e.g.~gradient-based methods) cannot reliably be used. Initially proposed by Glowinski et al.~\cite{fortin2000augmented,glowinski2011numerical}, the augmented Lagrangian method has found its own place as a robust and accurate method for solving these kinds of problems without regularizing the viscoplastic transition, or incorporating elastoviscoplasticity \cite{chaparian2019adaptive}. As shown by Frigaard and Nouar \cite{frigaard2005usage}, solving viscoplastic problems with unregularized rheology is vital to avoid computation of non-physical solutions when approaching the yield limit. Detailed reviews on utilizing this class of numerical methods in viscoplastic flow problems can be found in \cite{chaparian2019adaptive,saramito2017progress}. We avoid repeating all of the details here; however, in brief, this method is based on convex optimization which takes the advantage of transforming a viscoplastic flow problem to a saddle-point problem by means of variational tools (i.e.~minimum and maximum principles \cite{Huilgol2015}) and then a proper convex optimization algorithm (e.g.~the Uzawa approach) is used to find the velocity, pressure, and stress fields \cite{roquet2003adaptive}. This will be discussed in the next subsection.

The 2D flow considered numerically in the present manuscript can be represented by the non-dimensional governing and constitutive equations
\begin{equation}\label{non-govern}
0 = - \boldsymbol{\nabla} p + \boldsymbol{\nabla} \boldsymbol{\cdot} \boldsymbol{\tau}~~\text{in}~ \Omega \setminus\bar{X},
\end{equation}
and,
\begin{equation}\label{non-const}
  \left\{
    \begin{array}{ll}
      \boldsymbol{\tau} = \left( \Vert \dot{\boldsymbol{\gamma}} \Vert^{n-1} + \displaystyle{\frac{\mathcal{B}}{\Vert \dot{\boldsymbol{\gamma}} \Vert}} \right) \dot{\boldsymbol{\gamma}} & \mbox{iff}\quad \Vert \boldsymbol{\tau} \Vert > \mathcal{B}, \\[2pt]
      \dot{\boldsymbol{\gamma}} = 0 & \mbox{iff}\quad \Vert \boldsymbol{\tau} \Vert \leqslant \mathcal{B},
  \end{array} \right.
\end{equation}
respectively. To non-dimensionalize these equations the linear rotational velocity of the vane, $\hat{R}_{\text{1}} \hat{\Omega}_1$, is used as the velocity scale and the `gap' width $\hat{d}$ as the length scale. Both the pressure and the stress tensor are scaled with the characteristic viscous stress $ \hat{K}(\hat{R}_1\hat{\Omega}_1/{\hat{d}})^n$. Hence, the Bingham number is $\mathcal{B} = \frac{\hat{\tau}_y}{\hat{K}} \left( \frac{\hat{d}}{\hat{R}_1 \hat{\Omega}_1} \right)^n$. The dimensional torque on the vane is given by,
\[
\hat{\mathcal{T}} = \hat{K}\frac{(\hat{R}_1 \hat{\Omega}_1)^n}{\hat{d}^{n-2}} \mathcal{T},
\]
where $\mathcal{T}$ can be computed directly by integrating the couple resulting from the dimensionless stress acting everywhere on the vane surface,
\begin{equation}\label{eq:torque}
\mathcal{T} = \int_{\partial X} \boldsymbol{r} \boldsymbol{\times} \left[ (-p \boldsymbol{1} + \boldsymbol{\tau}) \boldsymbol{\cdot} \boldsymbol{n} \right] ~\text{d}S,
\end{equation}
where $\boldsymbol{r}$ is the position vector to a point on the vane surface and $\boldsymbol{1}$ is the identity tensor. We note that the torque can  alternatively be computed by integrating over the cup surface ($\partial \Omega$) since the flow is inertialess. In the case when the no-slip condition applies on the vane surface, the torque in this inertia-free creeping flow can also be computed via the energy balance equation \footnote{ The dimensional form of the energy balance equation reads
\[
\int_{\Omega \setminus \bar{X}} \hat{\boldsymbol{\tau}} \boldsymbol{:} \hat{\dot{\boldsymbol{\gamma}}} ~\text{d}\hat{A} = \int_{\Omega \setminus \bar{X}} \left[ \left( \hat{K} \Vert \hat{\dot{\boldsymbol{\gamma}}} \Vert^{n-1} + \displaystyle{\frac{\hat{\tau}_y}{\Vert \hat{\dot{\boldsymbol{\gamma}}} \Vert}} \right) \hat{\dot{\boldsymbol{\gamma}}} \right] \boldsymbol{:} \hat{\dot{\boldsymbol{\gamma}}} ~\text{d}\hat{A} =  \hat{\mathcal{T}} ~\hat{\Omega}_1
\]
or,
\[
\hat{K} \int_{\Omega \setminus \bar{X}} \Vert \hat{\dot{\boldsymbol{\gamma}}} \Vert^{n+1} ~\text{d}\hat{A} + \hat{\tau}_y \int_{\Omega \setminus \bar{X}} \Vert \hat{\dot{\boldsymbol{\gamma}}} \Vert  ~\text{d}\hat{A} = \hat{\mathcal{T}} ~\hat{\Omega}_1.
\]
Using the scales discussed earlier (i.e.~selecting the gap width $\hat{d}$ as the relevant length scale and $\hat{R}_1 \hat{\Omega}_1$ as the velocity scale), the non-dimensional form reduces to Eq.~\ref{eq:non_d_energy}.
},
\begin{equation}\label{eq:non_d_energy}
\int_{\Omega \setminus \bar{X}} \Vert \dot{\boldsymbol{\gamma}} \Vert^{n+1} ~\text{d}A + \mathcal{B} \int_{\Omega \setminus \bar{X}} \Vert \dot{\boldsymbol{\gamma}} \Vert  ~\text{d}A = \mathcal{T} \Big( \frac{1 - R}{R}\Big),
\end{equation}
where $R=\hat{R}_1 / \hat{R}_2$ is the radius ratio. The energy balance equation is still valid in the wall-slip regime but some extra terms appear on the left hand side of Eq.~\ref{eq:non_d_energy}. Please refer to \cite{chaparian2020sliding} for more details of this regime.

\begin{figure}[!h]
    \centering
    \includegraphics[width=0.8\textwidth]{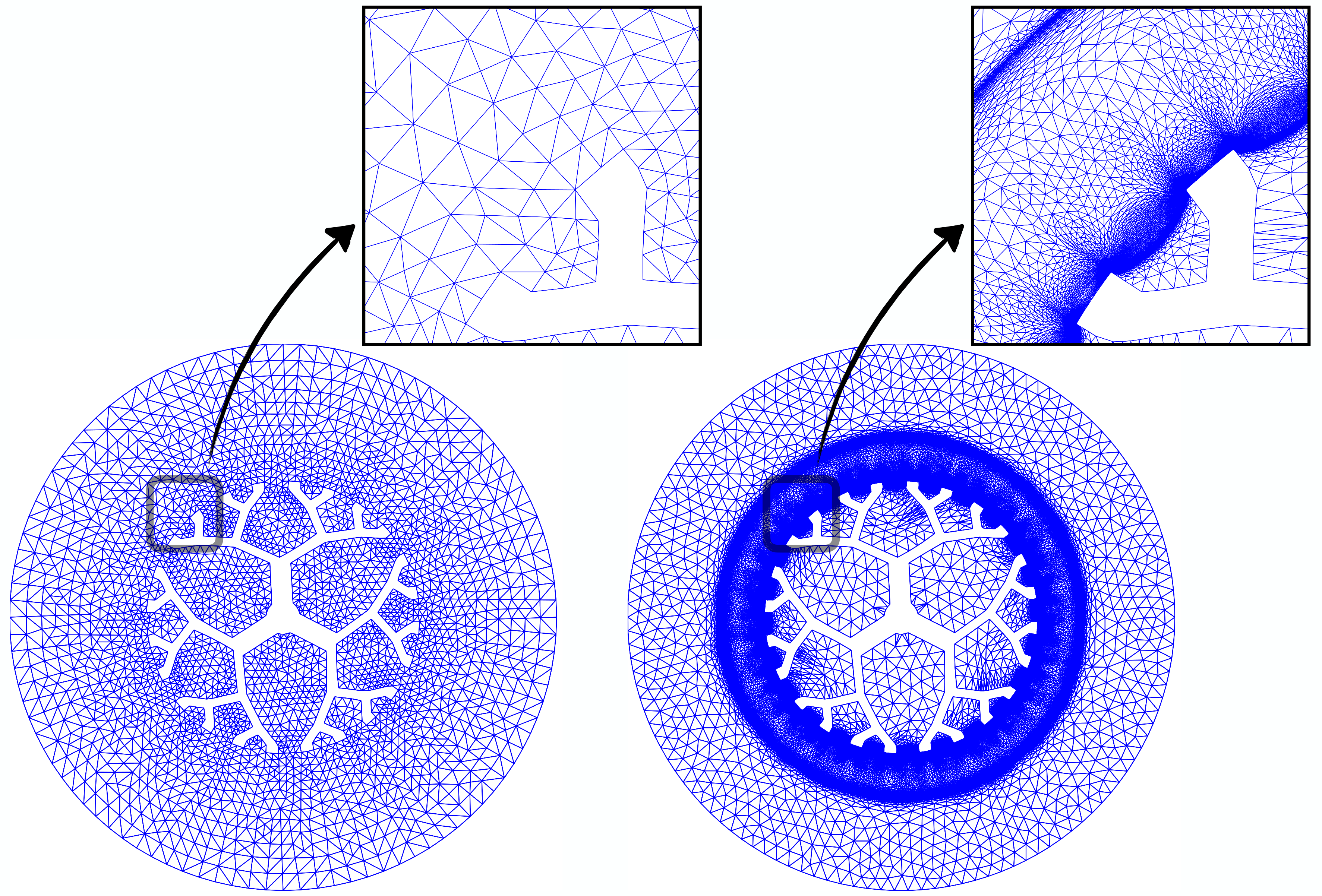}
    \caption{Adaptive finite element mesh refinement: left panel shows the initial mesh around a 24-arm fractal vane and the right panel shows the adapted mesh after eight cycles of refinement for a specific flow case.}
    \label{mesh}
\end{figure}

Regarding the discretization and meshing, we use the Finite Element Method (FEM) coupled with an anisotropic adaptive meshing algorithm \cite{roquet2003adaptive}. This guarantees sharper resolution of the yield surfaces, since the mesh elements are stretched anisotropically in the direction of the eigenvectors of the Hessian matrix of the dissipation function \cite{roquet2003adaptive}. In other words, elements are aligned/stretched anisotropically with the orientation of the local yield surfaces and also get locally refined to capture spatial variations in yield surfaces with a higher resolution. A sample meshing procedure is presented in Fig.~\ref{mesh}: the left panel shows the initial mesh and the right panel after eight cycles of adaptation. The implementation of the discussed numerical method and the mesh adaptivity is handled by an open-source library FreeFEM++ \cite{hecht2012}.

\subsection{Remarks on the presence of slip}

Since part of the current study deals with the hydrodynamic consequences of local slip over vane/cup surfaces, in this subsection, we briefly discuss the fluid slip law used in the present study and a general Uzawa algorithm which handles additional terms due to the slip. For more details, readers are referred to \cite{chaparian2020sliding}.

The general dimensional slip law for yield-stress fluids reads:
\begin{equation}
\hat{u}_{\text{s}} = \hat{\boldsymbol{u}}_{\text{ns}} \boldsymbol{\cdot} \boldsymbol{t} - \boldsymbol{\delta} \hat{\boldsymbol{u}} \boldsymbol{\cdot} \boldsymbol{t} = \left\{
\begin{array}{ll}
\hat{\beta}_s ~ \text{sgn}(\hat{\Lambda}) ~\left( \vert \hat{\Lambda} \vert - \hat{\tau}_s \right)^{\kappa}, & \text{iff}~~ \vert \hat{\Lambda} \vert > \hat{\tau}_s, \\[2pt]
0, & \text{iff}~~ \vert \hat{\Lambda} \vert \leqslant \hat{\tau}_s,
\end{array} \right.
\label{Sliplaw}
\end{equation}
where $\hat{u}_{{s}}$ is the tangential slip velocity on the solid surface $\partial \Pi$ ($\partial X$ and/or $\partial \Omega$), $\hat{\boldsymbol{u}}_{{ns}}$ the velocity of the solid boundary, and $\boldsymbol{\delta} \hat{\boldsymbol{u}}$ is the restriction of $\hat{\boldsymbol{u}}$ on $\partial \Pi$: $\hat{\boldsymbol{u}} \to \boldsymbol{\delta} \hat{\boldsymbol{u}}$ as $\hat{\boldsymbol{x}} \to \partial \Pi$. The critical value of the shear stress for onset of slip is designated by $\hat{\tau}_s$. The value of the tangential {\it traction vector} (i.e.~the tangential force per unit area acting on the solid surface) is given by $\hat{\Lambda} = \left[ \left( - \hat{p} \boldsymbol{1} + \hat{\boldsymbol{\tau}} \right) \boldsymbol{\cdot} \boldsymbol{n} \right] \boldsymbol{\cdot} \boldsymbol{t}$, where the normal and tangential unit vectors to the solid surface $\partial \Pi$ at each point are represented by $\boldsymbol{n}$ and $\boldsymbol{t}$, respectively. The no-penetration condition on $\partial X$ is enforced as well, so that $\boldsymbol{\delta}\hat{\boldsymbol{u}} \boldsymbol{\cdot} \boldsymbol{n} = \hat{\boldsymbol{u}}_{{ns}} \boldsymbol{\cdot} \boldsymbol{n} $. We note that for a general power-law relationship of the form of Eq.~(\ref{Sliplaw}) the physical dimensions of $\hat{\beta}_s$ are $m/(Pa^{\kappa} \cdot s)$. Hence, the relevant scale for the slip coefficient is $\displaystyle \frac{\hat{d}^{n\kappa}}{\hat{K}^{\kappa} (\hat{R}_1 \hat{\Omega}_1)^{n\kappa-1}}$. In the present study, we use the expression reported by Medina-Ba{\~{n}}uelos et al.~\cite{Medina-Banuelos2019},
\begin{equation}\label{eq:SlipMedina}
\hat{u}_s = \hat{\beta}_s ~\vert \hat{\Lambda} \vert^{1.8}
\end{equation}
with $\hat{\beta}_s = 1.04 \times 10^{-5}~m/(Pa^{1.8} \cdot s)$, unless stated otherwise, to make direct comparison possible with their measurements. In the presence of slip in the simplified form of Eq.~\ref{eq:SlipMedina} (i.e.~with no critical stress for onset of slip, $\hat{\tau}_s = 0$), the Uzawa algorithm takes the form of Algorithm 1 (displayed on the next page).

\begin{algorithm}[!h]
\caption{}\label{alg2}
\begin{algorithmic}[1]
\Procedure{}{solving yield-stress fluid flow with slip boundary condition}
\State $m \gets 0$
\vspace{3pt}
\State $ \boldsymbol{q}^0,\boldsymbol{\Xi}^0,\boldsymbol{\lambda}^0,\boldsymbol{\xi}^0 \gets \boldsymbol{0}~\text{(or any other initial guess)}$
\State \hspace{-12pt}{\it loop (Uzawa algorithm)}:
\If {$\text{residual} < \text{convergence}$} \textbf{close}.
\EndIf
\State find $\boldsymbol{u}^{m+1}$ and $p^{m+1}$ which satisfies $\forall (\boldsymbol{v}, \zeta)$,

$
\left\{
\begin{array}{l}
a \displaystyle \int_{\Omega \setminus \bar{X}} \dot{\boldsymbol{\gamma}} \left( \boldsymbol{u}^{m+1} \right) \boldsymbol{:} \boldsymbol{\nabla} \boldsymbol{v} ~\text{d}A - \int_{\Omega \setminus \bar{X}} p^{m+1} \left( \boldsymbol{\nabla} \boldsymbol{\cdot} \boldsymbol{v} \right) ~\text{d}A + \displaystyle \int_{\Omega \setminus \bar{X}} \left( \boldsymbol{\Xi}^m - a \boldsymbol{q}^m \right) \boldsymbol{:} \boldsymbol{\nabla} \boldsymbol{v} ~\text{d}A = \\[2pt]
- a \displaystyle \int_{\partial X} \boldsymbol{u}^{m+1} \boldsymbol{\cdot} \boldsymbol{v} ~\text{d}s + a \int_{\partial X} \boldsymbol{\xi}^m \boldsymbol{\cdot} \boldsymbol{v} ~\text{d}s + \int_{\partial X} \boldsymbol{\lambda}^m \boldsymbol{\cdot} \boldsymbol{v} ~\text{d}s, \\[2pt]
 \\[2pt]
\displaystyle \int_{\Omega \setminus \bar{X}} \zeta \left( \boldsymbol{\nabla} \boldsymbol{\cdot} \boldsymbol{u}^{m+1} \right)~\text{d}A = 0,
\end{array} \right.
$

with given no-slip B.C. (if there is any).

\vspace{5pt}
\State $\left\{
\begin{array}{ll}
~~~~~~~~~~~~~~~~~~~~~~~ \boldsymbol{q}^{m+1} \leftarrow 0, & \text{iff}~~ \Vert \boldsymbol{\Sigma} \Vert \leqslant B, \\[2pt]
\text{solve}  ~\boldsymbol{q}^{m+1} = \left( 1- \displaystyle\frac{B}{\Vert \boldsymbol{\Sigma} \Vert} \right) \displaystyle\frac{\boldsymbol{\Sigma}}{ \Vert \boldsymbol{q}^{m+1} \Vert^{n-1}+a} ~\text{for} ~\boldsymbol{q}^{m+1}, & \text{iff}~~ \Vert \boldsymbol{\Sigma} \Vert > B.
\end{array} \right.
$

\vspace{5pt} where $\boldsymbol{\Sigma} = \boldsymbol{\Xi}^m + a \dot{\boldsymbol{\gamma}}(\boldsymbol{u}^{m+1}) $

\vspace{5pt}
\State $\boldsymbol{\xi}^{m+1} \gets \displaystyle \left( \boldsymbol{u}_{\text{ns}} \boldsymbol{\cdot} \boldsymbol{n} \right) \boldsymbol{n} + \left( \boldsymbol{u}_{\text{ns}} \boldsymbol{\cdot} \boldsymbol{t} \right) \boldsymbol{t} + \frac{\beta_s}{1+\beta_s a} ~\Gamma~\boldsymbol{t}$

\vspace{5pt} where $\Gamma =  - \left( \boldsymbol{\lambda}^m \boldsymbol{\cdot} \boldsymbol{t} \right) \cdot \vert \boldsymbol{\lambda}^m \boldsymbol{\cdot} \boldsymbol{t} \vert ^{\kappa-1} + a~ \left( \boldsymbol{\delta u}^{m+1} \boldsymbol{\cdot} \boldsymbol{t} - \boldsymbol{u}_{\text{ns}} \boldsymbol{\cdot} \boldsymbol{t} \right)  $

\vspace{5pt}
\State $\boldsymbol{\Xi}^{m+1} \gets \boldsymbol{\Xi}^m + a \left[ \dot{\boldsymbol{\gamma}} \left(\boldsymbol{u}^{m+1}\right) - \boldsymbol{q}^{m+1} \right]$

\vspace{5pt}
\State $\boldsymbol{\lambda}^{m+1} \gets \boldsymbol{\lambda}^m - a \left[ \boldsymbol{\delta u}^{m+1} - \boldsymbol{\xi}^{m+1} \right]$

\vspace{5pt}
\State residual $\gets$
\begin{eqnarray*}
\max \biggl( \displaystyle \int_{\Omega} \vert \boldsymbol{u}^{m+1} - \boldsymbol{u}^m \vert &\text{d}A&, \displaystyle \int_{\Omega} \Vert \dot{\boldsymbol{\gamma}} \left(\boldsymbol{u}^{m+1}\right) - \boldsymbol{q}^{m+1} \Vert ~\text{d}A , \nonumber \\
\displaystyle \int_{\Omega} \Vert \boldsymbol{q}^{m+1} - \boldsymbol{q}^{m} \Vert &\text{d}A&, \int_{\partial X} \vert \boldsymbol{\lambda}^{m+1} - \boldsymbol{\lambda}^m \vert ~\text{d}S \biggl)
\end{eqnarray*}
\State $m \gets m+1$
\State \hspace{-12pt}\textbf{goto} \emph{loop}
\EndProcedure
\end{algorithmic}
\end{algorithm}

Upon convergence of algorithm \ref{alg2} with the free augmentation parameter $a$, the Lagrange multiplier $\boldsymbol{\Xi}$ converges to the {\it true} stress field, $\boldsymbol{q}$ to the {\it true} rate of strain tensor, the Lagrange multiplier $\boldsymbol{\lambda}$ to the traction vector on $\partial \Pi$, and the auxiliary variable $\boldsymbol{\xi}$ to the velocity on $\partial \Pi$. The full implementation of this algorithm has been extensively validated in the previous studies \cite{chaparian2020sliding,chaparian2017yield,chaparian2019porous}.

Regarding the computational domain, we mostly simulate the flow in the entire two-dimensional fluid domain since imposing the azimuthal symmetry boundary condition in a fraction of the domain is non-trivial in a Cartesian form for complex geometries e.g.~in section \ref{sec:FractalVane}. However, when   geometric simplification allows, we take advantage of imposing azimuthal symmetry to accelerate the simulations; e.g.~in Fig.~\ref{fig2} where only half of the entire gap is simulated.

\section{Simulation results}
Steady two-dimensional viscoplastic simulations were used to understand the effect of vane geometry and Bingham number $\mathcal{B}$ on the velocity field around the vanes, as well as the effect of the number of vane arms on deviations from a ``homogeneous shearing" or ``viscometric" flow. Slip was incorporated into additional simulations to understand the effect of wall slip on published experimental data. 

\subsection{Velocity fields around standard vanes}
\begin{figure}[!h]
    \centering
    \makebox[\textwidth][c]{\includegraphics[width=0.4\textwidth]{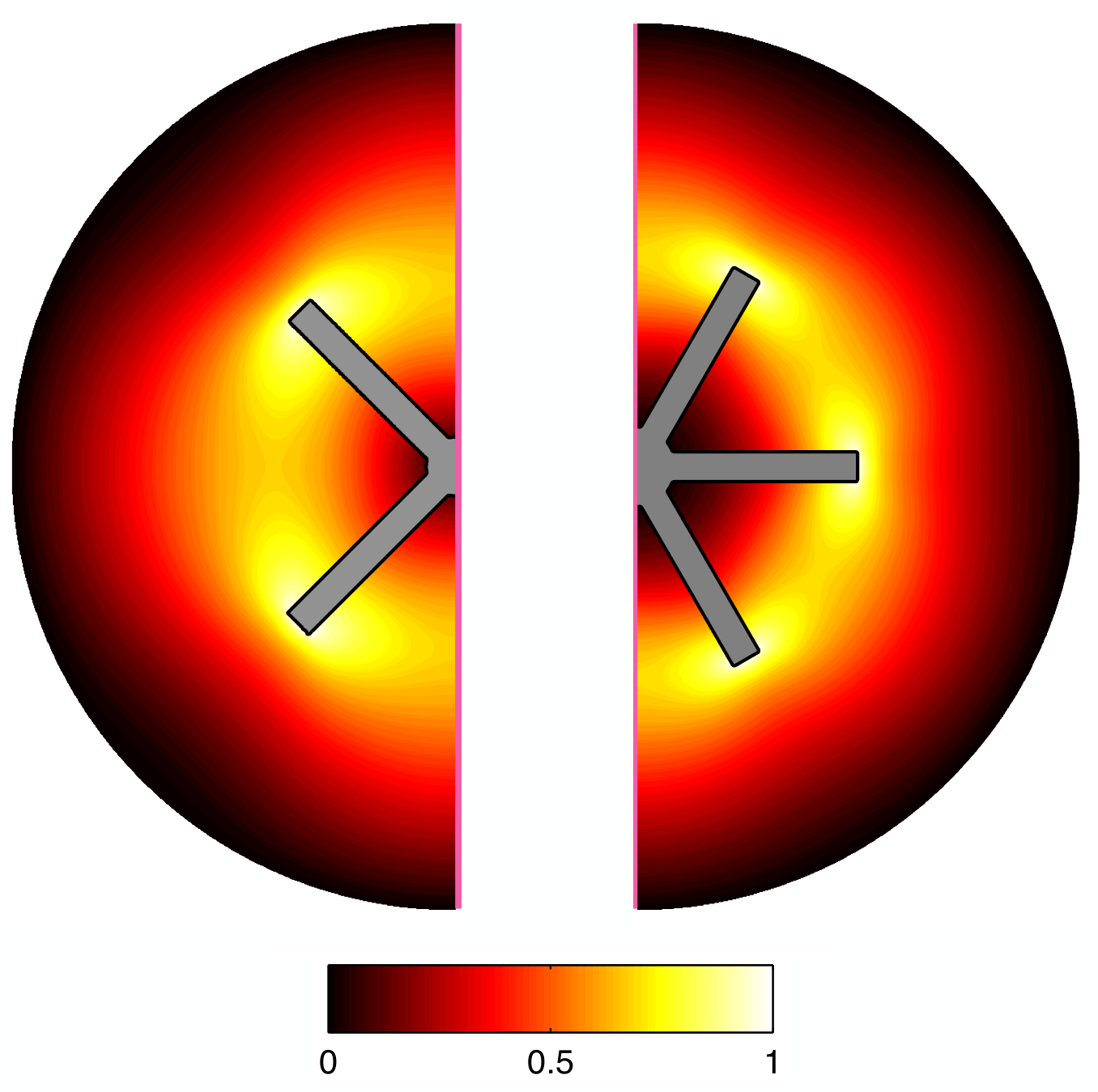}}%
    \caption{Features of viscous Newtonian flow around a 4-arm and 6-arm vane in a cup with $\hat{R}_2/\hat{R}_1=2$. The color map shows the dimensionless azimuthal velocity $\hat{v}_{\theta}/\hat{R}_1 \hat{\Omega}_1$ at $\mathcal{B} = 0$.}
    \label{fig2}
\end{figure}

The velocity field was first evaluated for the most common vane geometries having $N=4$ and $6$ at $\mathcal{B}=0$ (Newtonian flow) without slip, as shown in Fig.~\ref{fig2}. An azimuthally-periodic dependence on angle is evident, with relatively high velocity nodes directly around each of the vane tips, and local reductions of the tangential velocity in the interstitial spaces between vane arms. 

\begin{figure}[!h]
    \centering
    \makebox[\textwidth][c]{\includegraphics[width=0.7\textwidth]{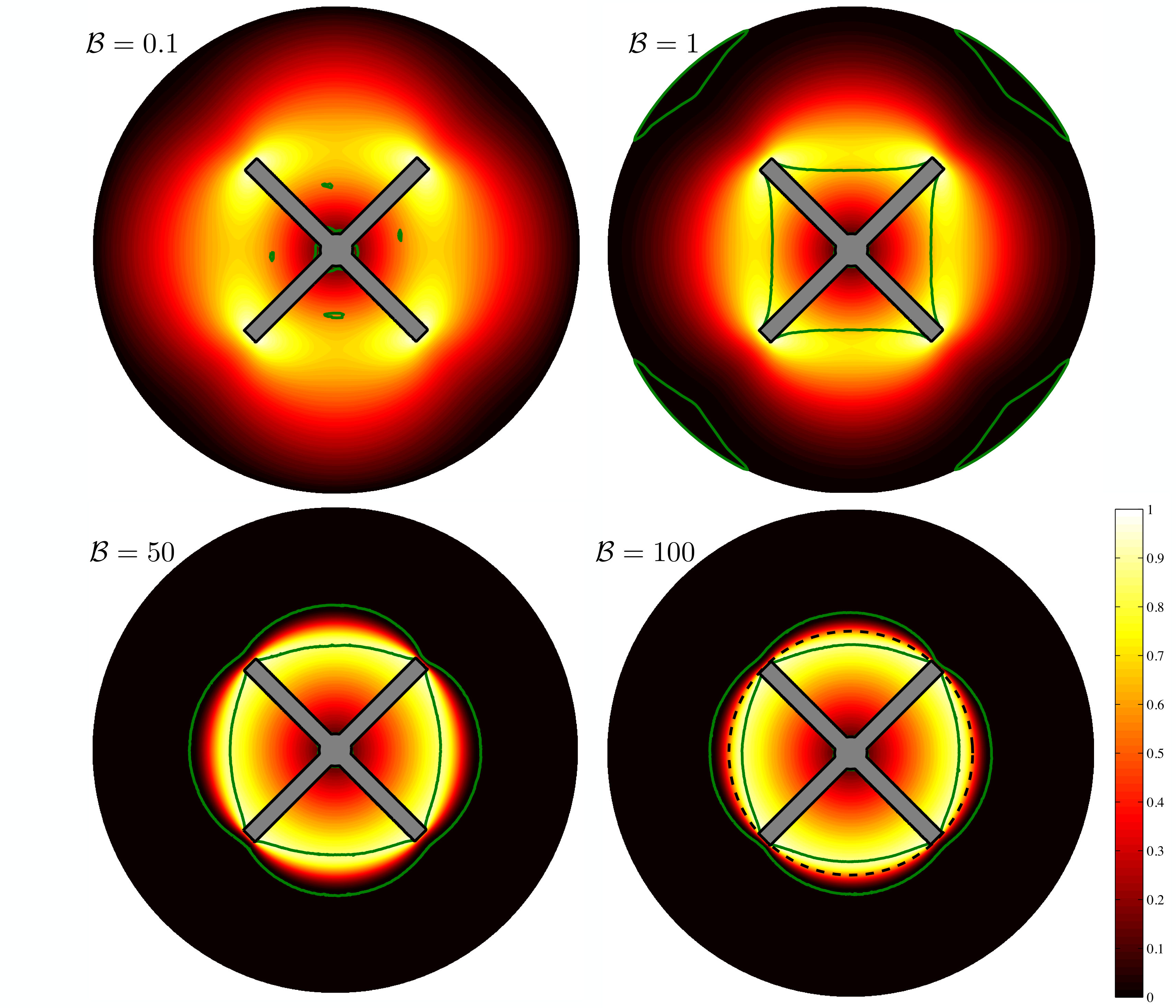}}%
    \caption{Colormap of the velocity magnitude $\vert \boldsymbol{u} \vert$ for different Bingham numbers. The green lines show the locations of the inner and outer yield surfaces and the dashed black circular line shows the extent of a circumscribed circle with radius $R_1$ for reference.}
    \label{fig4}
\end{figure}

In an axisymmetric wide-gap Taylor-Couette geometry the local tangential stress decreases as $\tau_{r \theta} \sim r^{-2}$. As the Bingham number of the flow is increased (either by increasing the yield stress of the fluid or by reducing $\hat{\Omega}$), yield surfaces emerge in the fluid at points in which the local shear stress equals the yield stress value, as highlighted by the green lines in Fig.~\ref{fig4} around the $4$-arm vane. A very weak secondary flow or ``Moffatt corner vortex'' exists between each pair of vane arms (which act effectively as the bounding walls of a lid-driven cavity) even for a Newtonian fluid \cite{Gutierrez-Barranco2013, Atkinson1992a}. For $\mathcal{B}=0.1$, the core at the center of each recirculation between the vane arms is unyielded, while the majority of the fluid is yielded throughout the entire domain. For $\mathcal{B}=1$, both inner and outer yield surfaces are visible. The outer yield surface is partially in contact with the stationary outer wall of the cup, but deviates radially inwards at four locations corresponding to the instantaneous positions of the vane arms, at which the local shear rate in the fluid gap is greatest. The inner yield surface outlines an approximately square plug of fluid that rotates as a solid body with the rigid cruciform vane structure. As $\mathcal{B}$ increases further, the inner yield surface stays pinned to the ends of the rotating vane arms and the outer yield surface moves radially inwards so that progressively less fluid is sheared. In the open regions between each arm the region of sheared fluid expands radially. Meanwhile, the flow retains a $4$-fold periodic structure for all $\mathcal{B}$, even for $\mathcal{B}=100$ where the inner and outer yield surfaces are nearly in contact at the vane tips, as shown in Fig.~\ref{fig4}(d).  

\begin{figure}[!h]
    \centering
    \makebox[\textwidth][c]{\includegraphics[width=0.75\textwidth]{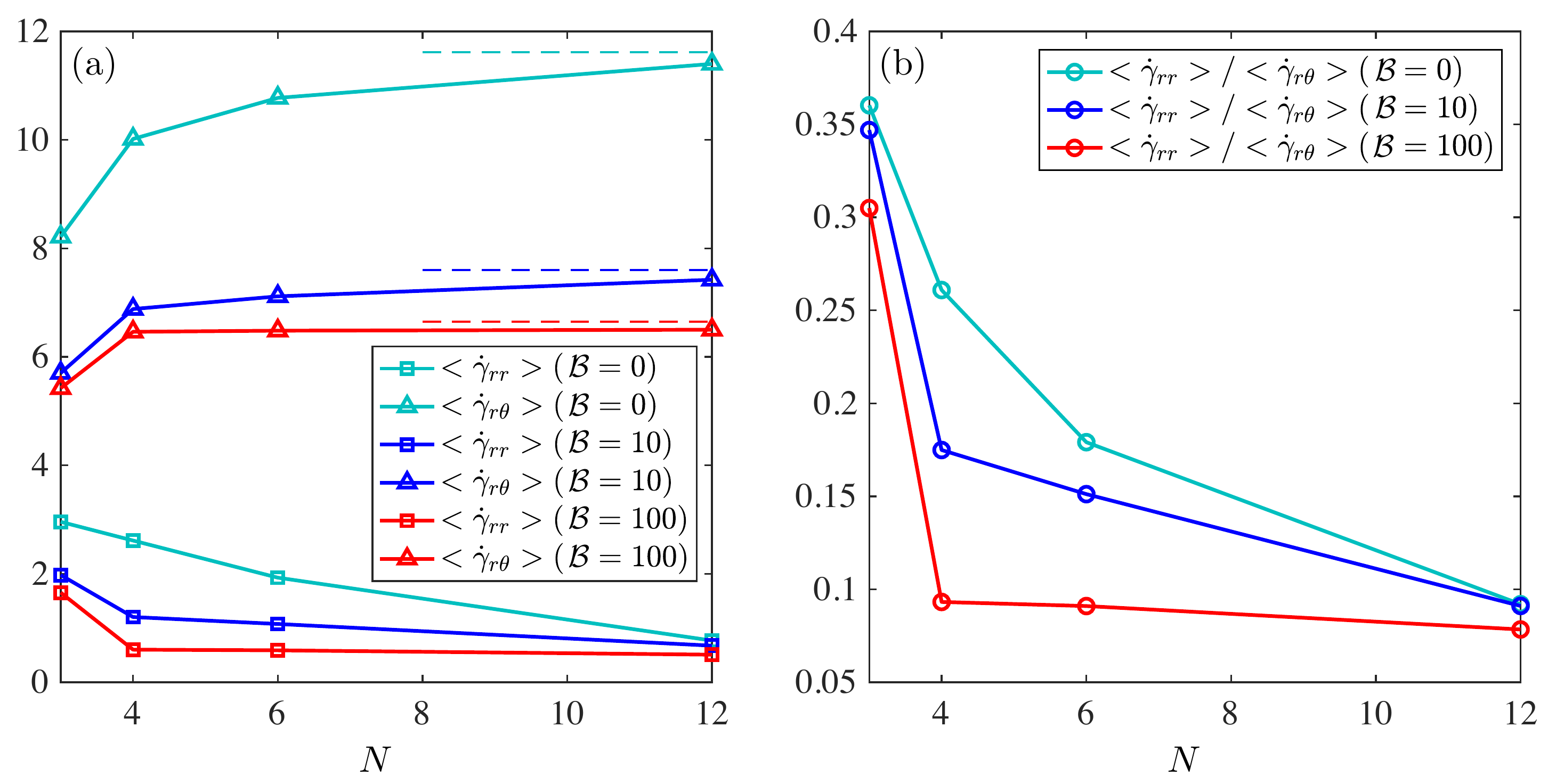}}%
    \caption{(a) $L^2$-norms ($<\cdot> = \int_{\Omega \setminus \bar{X}} \cdot ~\text{d}A$) of the variation in the extensional component $\dot{\gamma}_{rr}$ and shearing component $\dot{\gamma}_{r\theta}$ of the strain rate tensor with respect to the number of arms of the vane and a range of different Bingham numbers, (b) the ratio of the $L^2$-norms shown in panel (a). Dashed lines in (a) indicate the limiting result for a uniform rigid cylindrical rotor with no slip.}
    \label{fig5-grr}
\end{figure}

When the velocity field has periodic azimuthal variations, the shear rate is no longer homogeneous, and the flow is no longer truly ``viscometric". For a complex fluid this will lead to systematic errors in the measurement of a steady shear viscosity from integral quantities such as the total torque acting on the vane arms.  This effect is quantified by the ``$L^2$-norm" ($<\cdot> = \int_{\Omega \setminus \bar{X}} \cdot ~\text{d}A$) of the local radial extension rate $\dot{\gamma}_{rr}$ and tangential shear rate $\dot{\gamma}_{r\theta}$ of the deformation rate tensor. We compute these components as well as the corresponding ratio over the entire fluid domain for a series of vanes with $N=3, 4, 6, 12$, and for an ideal cylinder over a range of Bingham numbers $\mathcal{B}=0, 10, 100$ as shown in Fig.~\ref{fig5-grr}. For $N=3$,  $\dot{\gamma}_{rr}/\dot{\gamma}_{r\theta} > 0.3$ for all $\mathcal{B}$, which deviates markedly from ideal simple shear flow. As $N$ increases, extensional deformation between the vanes $\dot{\gamma}_{rr}$ decreases and the local homogeneity of the shear rate $\dot{\gamma}_{r\theta}$ increases. When $N=12$, the ratio shown in Fig.~\ref{fig5-grr}(b) falls to $\dot{\gamma}_{rr}/\dot{\gamma}_{r\theta}<0.1$ for all $\mathcal{B}$, indicating a close approximation of simple shear flow. For $N=12$, the calculated value of $\dot{\gamma}_{r\theta}$ approximately equals the dimensionless shear rate calculated for a cylinder under the same flow condition; see \ref{sec:Couette} for the analytical expression reproduced from \cite{Landry2006,chaparian2019adaptive}. We note that local contributions from the transition to a no-slip BC at the tip of each vane arm prevents $\dot{\gamma}_{rr}$ from reaching zero for all finite $N$. As the Bingham number increases, 
 the ratio of $\dot{\gamma}_{rr}$/$\dot{\gamma}_{r\theta}$ also decreases monotonically indicating that the flow becomes increasingly viscometric in character at higher Bingham numbers (i.e.~as we approach yield limit) and for vanes with more arms. 

\subsection{Effect of wall slip on velocity profiles} 

\begin{figure}[!h]
    \centering
    \makebox[\textwidth][c]{\includegraphics[width=0.8\textwidth]{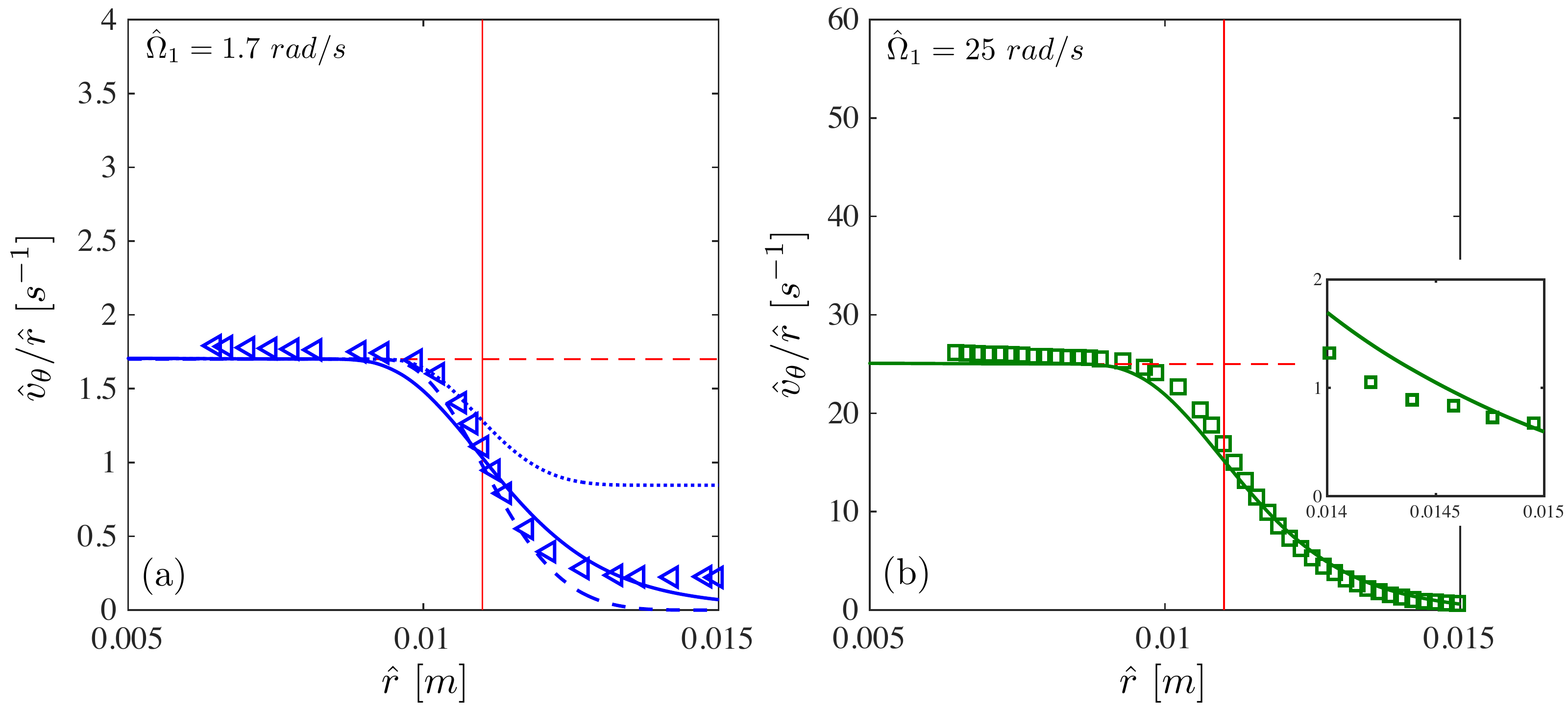}}%
    \caption{Angular velocity ($\hat{v}_{\theta} / \hat{r}$) profiles for the configuration of Medina-Ba{\~{n}}uelos et al.~\cite{Medina-Banuelos2019} at different rotation speeds; the continuous lines in both panels represent computations with the slip law Eq.~(\ref{eq:SlipMedina}) (i.e.~$\hat{\beta}_s = 1.04 \times 10^{-5}~m/(Pa^{1.8} \cdot s)$ and $\kappa=1.8$) and the symbols are experimental data from \cite{Medina-Banuelos2019}: (a) $\hat{\Omega}_1 = 1.7 ~rad/s$, and (b) $\hat{\Omega}_1 = 25 ~rad/s$; the inset of this panel shows an expanded view close to the cup wall. In the panel (a) two additional computations are shown for the sake of comparison with the continuous blue line: the dotted blue line illustrates the effect of changing the slip coefficient to $\hat{\beta}_s = 5.27 \times 10^{-5}~m/(Pa^{1.8} \cdot s)$ and the dashed blue line illustrates the expected velocity profile with the no-slip condition enforced at the outer wall. The red lines are  markers delineating the  edge of the geometry and the imposed vane angular velocity: continuous red line is the outer edge of each vane arm (i.e.~$\hat{R}_1$), and the dashed red line shows the imposed angular velocity of the vane fixture, $\hat{\Omega}_1$.}
    \label{fig7}
\end{figure}

We also compared the simulations with experimental measurements described by Medina-Ba{\~{n}}uelos et al.~\cite{Medina-Banuelos2019}, in which a $6$-arm vane was used to measure the rheological response of Carbopol in a cup with $\hat{R}_2/\hat{R}_1=1.4$. This study included bulk rheometric data as well as slip and velocity fields measured using PIV. Our simulations are designed to explore the effects of wall slip on viscometric experiments with vane fixtures. Three conditions are explored: no slip, the slip law of Eq.~(\ref{eq:SlipMedina}) having $\hat{\beta}_s = 1.04 \times 10^{-5}~m/(Pa^{1.8} \cdot s)$, and the same slip law with an increased value of the coefficient $\hat{\beta}_s = 5.27 \times 10^{-5}~m/(Pa^{1.8} \cdot s)$ to understand the sensitivity of the profile to the slip coefficient. The results are shown in Fig.~\ref{fig7} in terms of the radial variation in the local angular velocity of fluid elements $\hat{\Omega}(\hat{r}) = \hat{v}_{\theta}/\hat{r}$. Within most of the cruciform vane structure ($\hat{r}/R_1 < 1$), the velocity grows linearly with radius, and the material is thus in rigid body rotation. However, as we approach the outer edge of the vane arm, the velocity begins to decrease markedly and differences are visible between the different slip conditions. At the lower velocity of $\hat{\Omega}_1 = 1.7 ~rad/s$,  large differences between the local azimuthal velocity near the outer wall are apparent for different slip conditions, and the velocity at $\hat{r} = \hat{R}_2$ reaches a non-zero value. At a higher rotation rate, $\hat{\Omega}_1=25$ rad/s, only the slip law of Eq.~(\ref{eq:SlipMedina}) is shown, and little-to-no slip is observed at the outer wall. Indeed, at high angular rotation rates (e.g.~25 $rad/s$), the slip velocity at the cup wall is negligible compared to the high velocity of the material adjacent to the vane and in the gap. However, when compared to the low rotation rates (e.g.~1.7 $rad/s$), the magnitude of the slip velocity is higher which is intuitive from Eq.~(\ref{eq:SlipMedina}): the larger the shear stress is at the solid boundary, the larger the slip velocity is.

\subsection{End effects}

The numerical simulations performed in this paper are two-dimensional and thus only incorporate the cross-sectional geometry of the vanes. Local PIV-derived velocity profiles can be directly compared between experiments and simulations, but detailed comparisons of integrated quantities such as torque are more delicate because of three dimensional end-effects. When a real vane geometry is inserted in a fluid and rotated the total torque is a sum of the moments arising from shear forces acting on the faces (and tips) of all of the vane arms  plus the two end-caps \cite{Nguyen1983}. The proportion of the total torque due to end effects in a physical experiment with a finite length vane can be estimated based on the torque-to-shear stress conversion given in Eq.~\ref{eq:converttorque} with, and without, the end effect contribution: 
\begin{equation}\label{eq:endeffects}
\frac{\mathcal{T}_{\text{\it end effect}}}{\mathcal{T}_{total}} = \frac{S_{N,2D}}{S_{N}} = \frac{\frac{\hat{R}_1}{4\hat{L}}(2.75-\frac{3}{\sqrt{N}})}{(1-\frac{1.113}{N})+\frac{\hat{R}_1}{4\hat{L}}(2.75-\frac{3}{\sqrt{N}})}
\end{equation}

\begin{figure}[!h]
    \centering
    \includegraphics[width=0.5\textwidth]{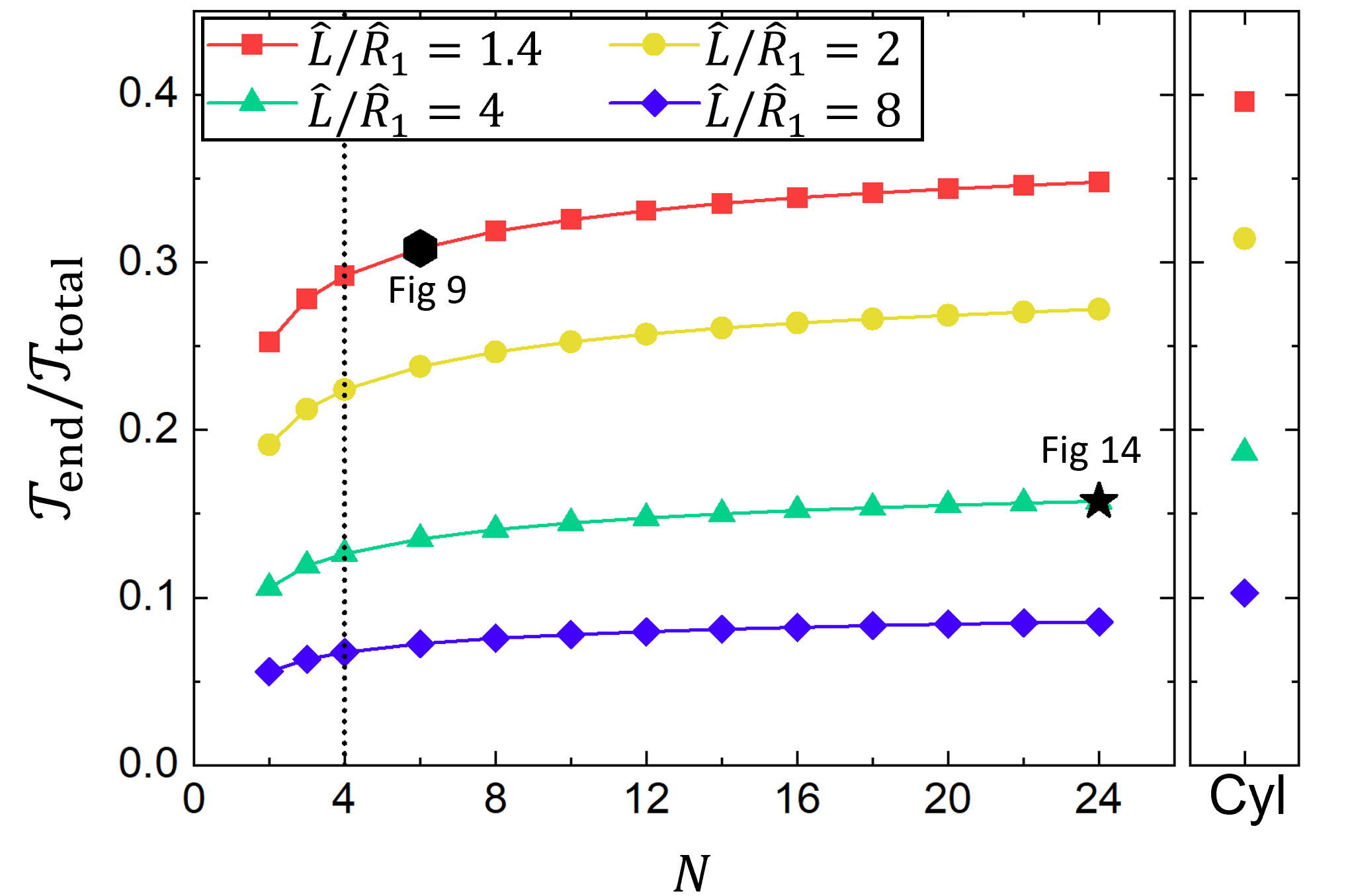}
    \caption{The proportion of total torque arising due to end effects as calculated from Eq.~\ref{eq:endeffects}. The contribution of torque due to the end caps of the vanes is reported as a function of the vane aspect ratio and the number of vane arms, with the values on the righthand plot showing the terminal values for cylinders of the corresponding aspect ratio. The values used to analyze experimental data in Figures~\ref{fig8} and \ref{fig:total_torque} are indicated on the plot.
    }
    \label{fig9}
\end{figure}

As a result of the weak  $N^{-1/2}$ decay with the number of arms, this contribution remains substantial, typically accounting for $10\%$ or more of the measured torque in most experiments, even if the vane is relatively slender (\textit{e.g,}, if $\hat{L}/\hat{R}_1=4$), as calculated explicitly in Fig.~\ref{fig9}. To minimize boundary effects, experimental protocols typically recommend submerging the vane to a depth such that the extent of the fluid between the bottom surface of the vane and the base of the cup is greater than or equal to the vane diameter, plus a corresponding distance between the top of the vane and the free surface of the fluid sample greater than or equal to the vane radius, as depicted graphically in Fig.~\ref{fig1} \cite{Macosko1994_TCintro}. If an insufficient lower spacing is set in experiments, the end effect contributions are augmented by a non-negligible contribution to the total measured torque due to torsional shearing flow between the vane and the horizontal cup surface, particularly if a wide-gap configuration is used \cite{Lindsley1947}. the value of the end effect ratio may increase by $0.03$ or more for substantially thixotropic fluids as well as for simple yield-stress fluids at very low shear rates \cite{Potanin2010}. In comparison, a DIN (53019) rotor bob maintains a radius ratio, $\hat{R}_2/\hat{R}_1$, near unity, giving a typical end effect contribution of about $11\%$ \cite{Giles2011}. For solid Couette bobs, this end effect may also be reduced by using a rotor with a recessed end-section that traps an air bubble \cite{Lindsley1947}, or by using stationary ring guards as end plates with the rotating cylinder located in between, as Maurice Couette himself first used \cite{Piau2005, Macosko1994_endeffects}. However, these methods are not compatible with vane geometries.

\begin{figure}[!h]
    \centering
    \makebox[\textwidth][c]{\includegraphics[width=0.8\textwidth]{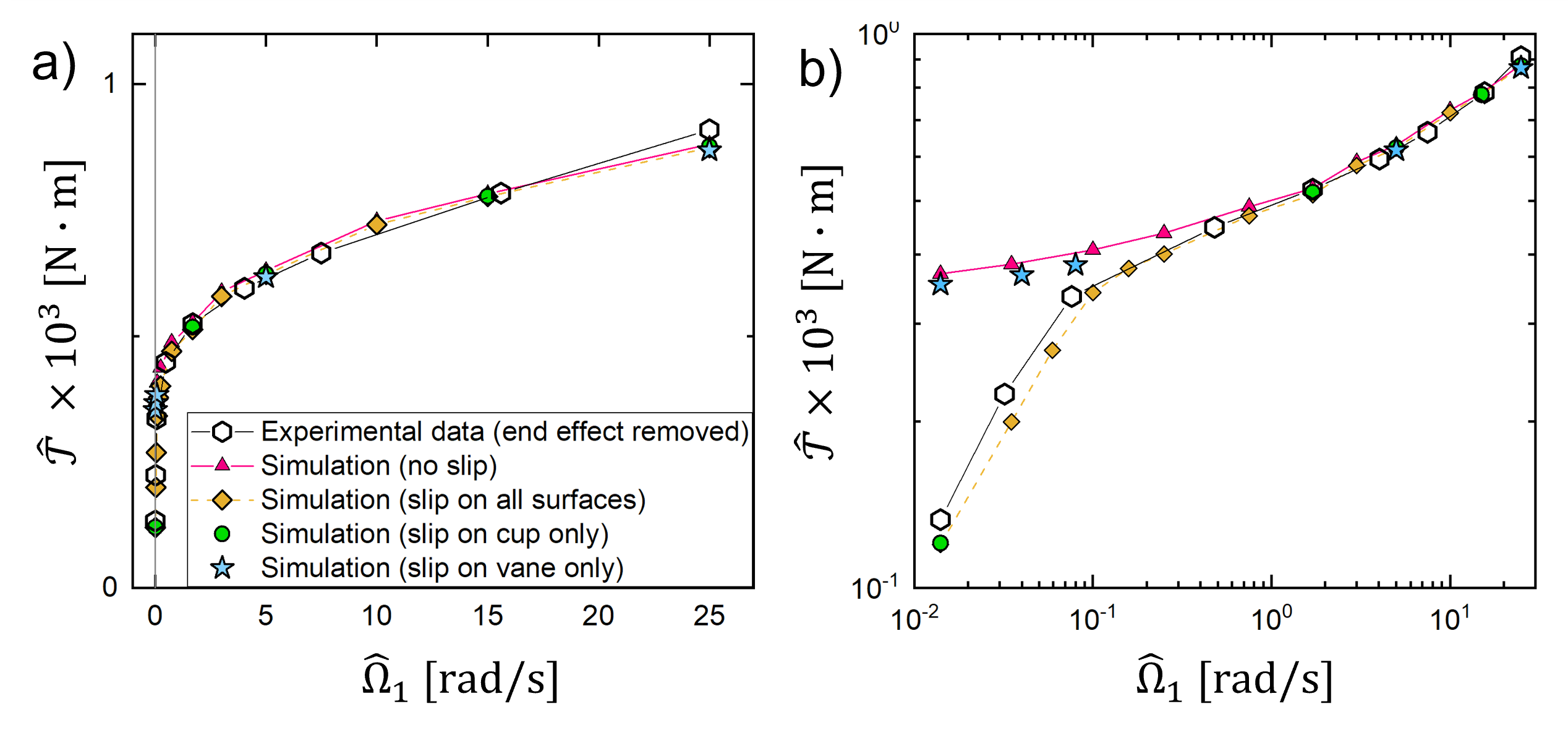}}%
    \caption{Computed torque on a 6-arm vane in a cup with $\hat{R}_2/\hat{R}_1=1.4$ compared to published results for a yield-stress fluid \cite{Medina-Banuelos2019} plotted on (a) linear and (b) logarithmic scales. The contribution from end effects present in the data for total torque were accounted for approximately using Eq.~\ref{eq:endeffects}. Simulations show the computed total torque for a cup having slip (yellow - - -) or no-slip ( red ----) boundary conditions on both the outer cup and vane, as well as the case of slip on the cup only (green circles) and slip on the vane only (blue stars). A slip boundary condition on the rotating vane surface does not perceptibly affect the total torque.}
    \label{fig8}
\end{figure}

For the $6$-arm vane, Eq.~(\ref{eq:torque}) was used to calculate the  total torque per unit length measured by the simulation over a range of shear rates, generating a flow curve which could be compared to the measured torque $\hat{\mathcal{T}}$ vs rotation rate  $\hat{\Omega}_1$ results from Medina-Ba{\~{n}}uelos et al.~\cite{Medina-Banuelos2019}. Using Eq.~\ref{eq:endeffects}, the experimentally-measured torque data was approximately separated into the torque from the cylindrical face plus the end effect contribution, which is typically about 30\%, of the total as shown in Fig.~\ref{fig8} for a vane having $\hat{L}/\hat{R}_1=1.4$ and $N=6$. To investigate the role of slip on the experimental data, four sets of simulations were performed using different surface slip conditions: (1) no slip, (2) slip on both the outer wall \textit{and} the vane, using the slip law of Eq.~\ref{eq:SlipMedina}, (3) slip on the outer wall only, and (4) slip on the vane only. The results are plotted alongside the experimental data in Fig.~\ref{fig8} wherein the two-dimensional torque data from the simulations was multiplied by the length of the experimental vane to convert to a true, three-dimensional torque. All simulation results agree well with the experimental data at high rotation rates $\hat{\Omega}_1>1$ rad/s. Notably, the effect of slip is still apparent in the velocity profiles presented in Fig.~\ref{fig7} for $\hat{\Omega}_1 = 1.7$ rad/s, whereas at this rotation rate the effect on the integrated torque is much smaller. It is only when plotted on a logarithmic abscissa that the effects of slip become apparent. In Fig.~\ref{fig8}(b) the experimental data at low rotation rates $\hat{\Omega}_1< 0.5$ rad/s, as well as the simulations incorporating slip, are clearly distinct from the reference simulation of case (1) with no slip. The two torque curves that are computed with wall slip (cases (2) and (3)) deviate systematically below the expected material torque measurement predicted from simulation case (1) without any wall slip and the case (4) with slip only on the vane. However, at all simulated rotation rates, the computed torque is nearly equal for cases (2) and (3), (i.e.~for slip only at the outer wall and for slip on both surfaces respectively), and both agree well with the observed experimental data. 

Two intriguing conclusions can be drawn from these results. The first is that the near-perfect agreement between the simulation with slip and with no-slip boundary conditions on the vane supports the assertion that a vane experiences minimal slip artifacts, even if the material it is constructed from (e.g.~a smoothly-machined metal) is prone to slip against a particular fluid sample. The cruciform geometry of the vane is able to overcome unfavorable interactions between the vane material surface and the fluid it is immersed in. However, secondly, it can be noted that even when using a non-slipping vane, if the outer wall of the cup containing the test fluid is liable to induce slip, then artefacts of this local failure in the no-slip boundary condition will remain present in the torque data, corrupting rotation rate measurements. Slip-free measurements thus require both a slip-free tool and a slip-free cup.  The slip documented in the experimental data of Medina-Ba{\~{n}}uelos et al.~at low rotation rates below $\hat{\Omega}_1 = 0.5$ rad/s, can therefore be ascribed principally to slip on the outer cup wall, as a vane geometry was utilized. 

\section{Fractal vane simulations}\label{sec:FractalVane}

\begin{figure}[!h]
    \centering
    \makebox[\textwidth][c]{\includegraphics[width=0.7\textwidth]{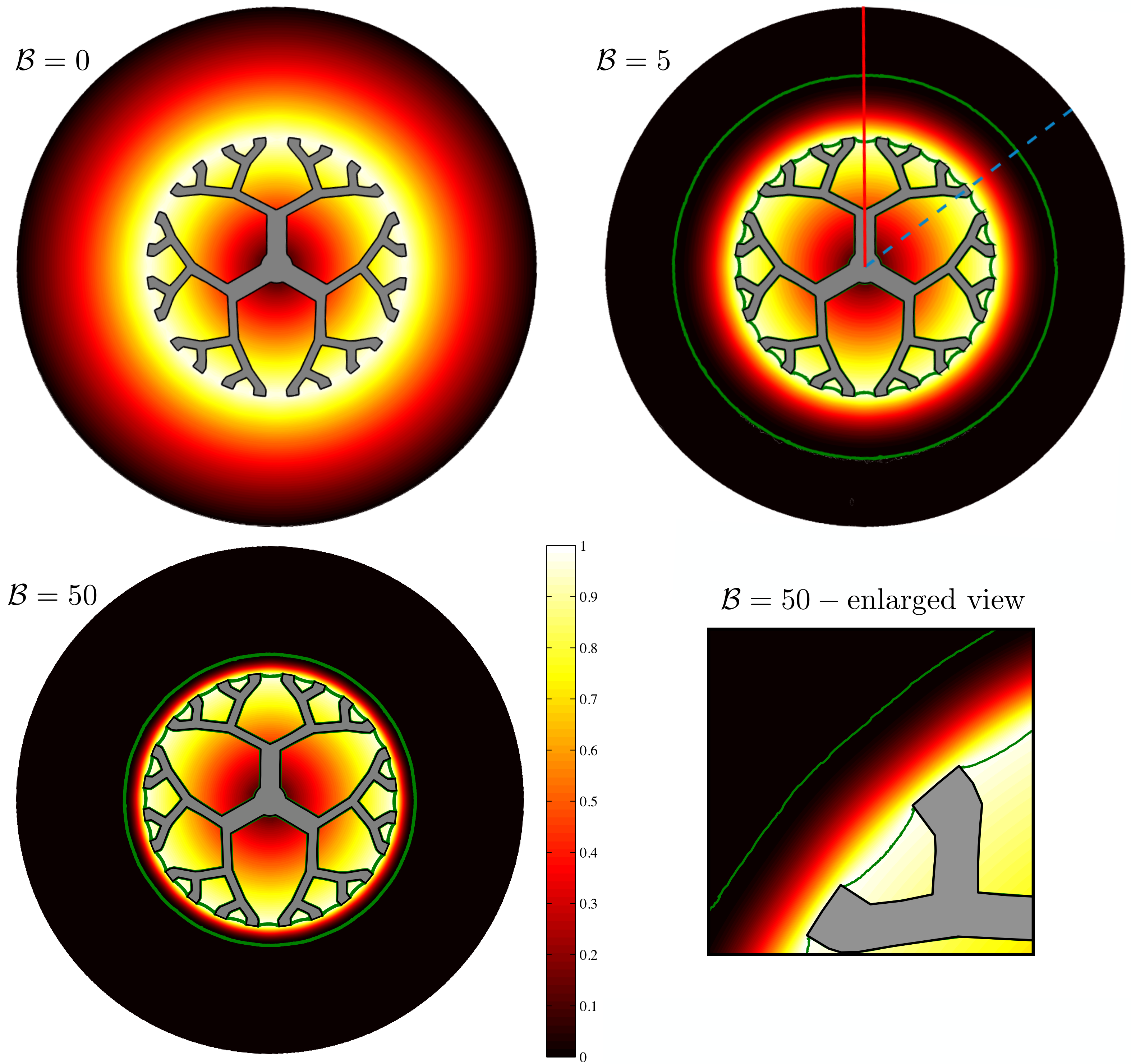}}%
    \caption{Colormap of the magnitude of the velocity fields ($\vert \boldsymbol{u} \vert$) around 24-arm fractal vanes for a range of Bingham numbers in a cup with $\hat{R}_2/\hat{R}_1=2$. Green lines indicate the inner and outer yield surfaces. The azimuthal velocity distribution along the red continuous and blue dashed lines in the top right panel are presented in Fig.~\ref{fig11-velocity} with the same color scheme.}
    \label{fig10-fractals}
\end{figure}

\begin{figure}[!h]
    \centering
    \makebox[\textwidth][c]{\includegraphics[width=0.9\textwidth]{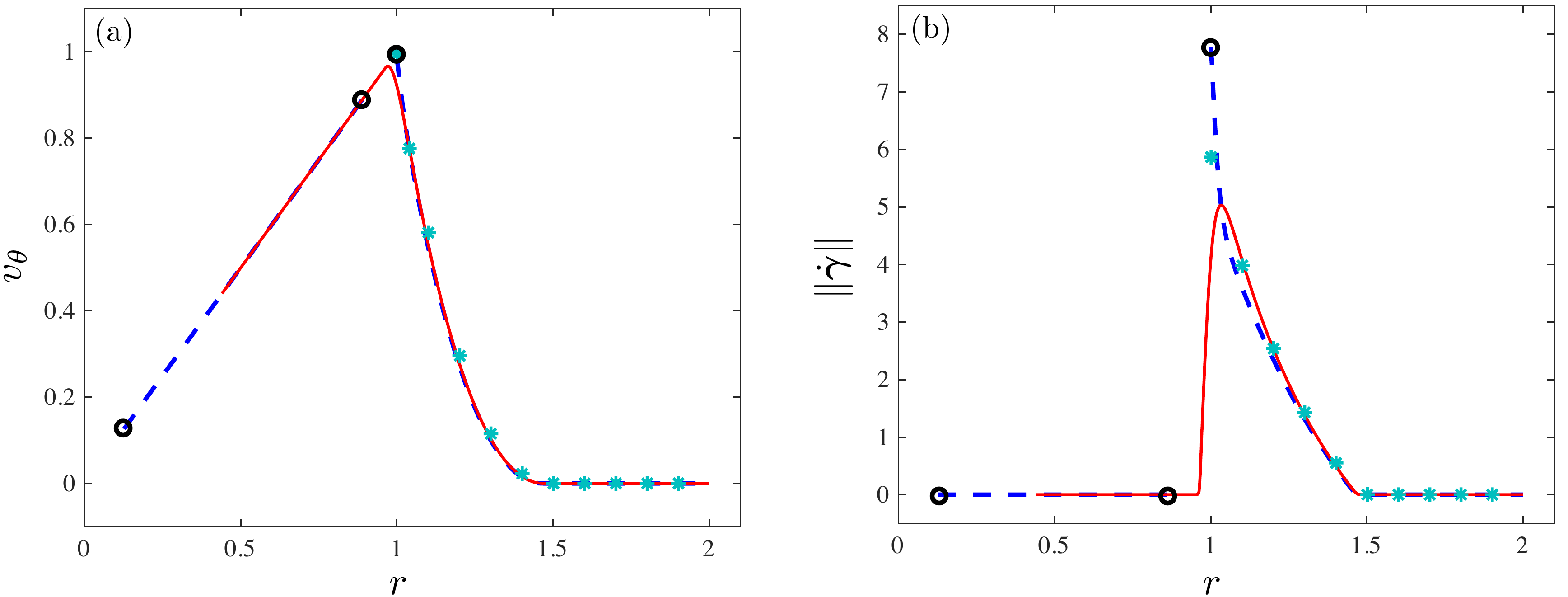}}%
    \caption{(a) Azimuthal velocity and (b) magnitude of the strain rate profiles along the red continuous and blue dashed lines in the top right panel of Fig.~\ref{fig10-fractals} ($\mathcal{B}=5$) with the same color interpretation. Please note that the part inside the fractal arm is not shown in the blue dashed line and the points on the vane surface are marked by the black circles. The cyan symbols show the analytical solution for simple one-dimensional axisymmetric Couette flow between two cylinders $\hat{R}_1 \le \hat{r} \le \hat{R}_2$; see \ref{sec:Couette}.}
    \label{fig11-velocity}
\end{figure}

We now explore how the kinematics are changed if fractal vane geometries such as those shown in Fig.~\ref{fig:Schematic} are used in place of the normal 4- or 6-arm cruciform design. The simulated velocity fields for Bingham fluids around $24$-arm fractal vanes and over a range of Bingham numbers are presented in Fig.~\ref{fig10-fractals}. Green outlines indicate the position of inner and outer yield surfaces. For $\mathcal{B}=0$, corresponding to a Newtonian fluid, the fluid is fully yielded and the flow is sensitive to the full shape of the vane, although the velocity profile has no distinguishable azimuthal structure at this scale. At high $Re$ we may expect to see some fluid recirculation within the vane arms (which each effectively act as a lid-driven cavity). At low $Re$ and $\mathcal{B}$ (i.e.~Newtonian limit), as is typical in $4$-arm vane rheometry, the fractal structure is expected to significantly weaken the strength of this secondary flow as it reduces the largest internal characteristic lengthscale. 
As the Bingham number is incremented to $\mathcal{B} = 5$ two yield surfaces appear. The outer yield surface is almost perfectly axisymmetric but some azimuthal variations in the locus of the inner yield surface can be detected from vane tip to tip. To illustrate this more directly, in Fig.~\ref{fig11-velocity}(a) we present radial line cuts of the azimuthal velocity profile for $\mathcal{B}=5$ through the opening between vane arms and through the solid vane structure as indicated by the solid and dashed radial lines in Fig.~\ref{fig10-fractals}. The two velocity profiles are indistinguishable for most radial positions, and within the vane structure, the fluid velocity increases linearly with radial position. At a radius just less than $\hat{r}=\hat{R}_1$, the fluid velocity begins to decrease in the opening between two adjacent vane arms; however, at larger $r$ the velocities along each line again correspond closely to each other. The corresponding profiles of the strain rate tensor (as represented by the second invariant) for the same two radial profiles are shown in Fig.~\ref{fig11-velocity}(b) and it is clear that the changes in the velocity profile lead to markedly different shear rates very close to the solid vane surface. 

As the Bingham number increases from $\mathcal{B}$ = 5 to $50$, the outer yield surface moves progressively closer to the surface of the vane, but remains approximately circular. The locus of the inner yield surface is also remarkably invariant, and remains pinned by the finely spaced solid perimeter surfaces of the fractal vane tips, as shown more clearly in the zoomed-in image in Fig.~\ref{fig10-fractals}. Fluid within the inner yield surface rotates as a solid, unyielded plug with the rigid vane rotor. As a result of this pinning, the $24$ contact points fix the location of the yield surface and effectively ``cloak" the inner structure of the fractal vane from the shear flow in the annulus,  so the velocity field is insensitive to the details of the fractal structure \cite{Chaparian2017}. 

\subsection{Viscoplastic boundary layer}
\begin{figure}[!h]
    \centering
    \makebox[\textwidth][c]{\includegraphics[width=1\textwidth]{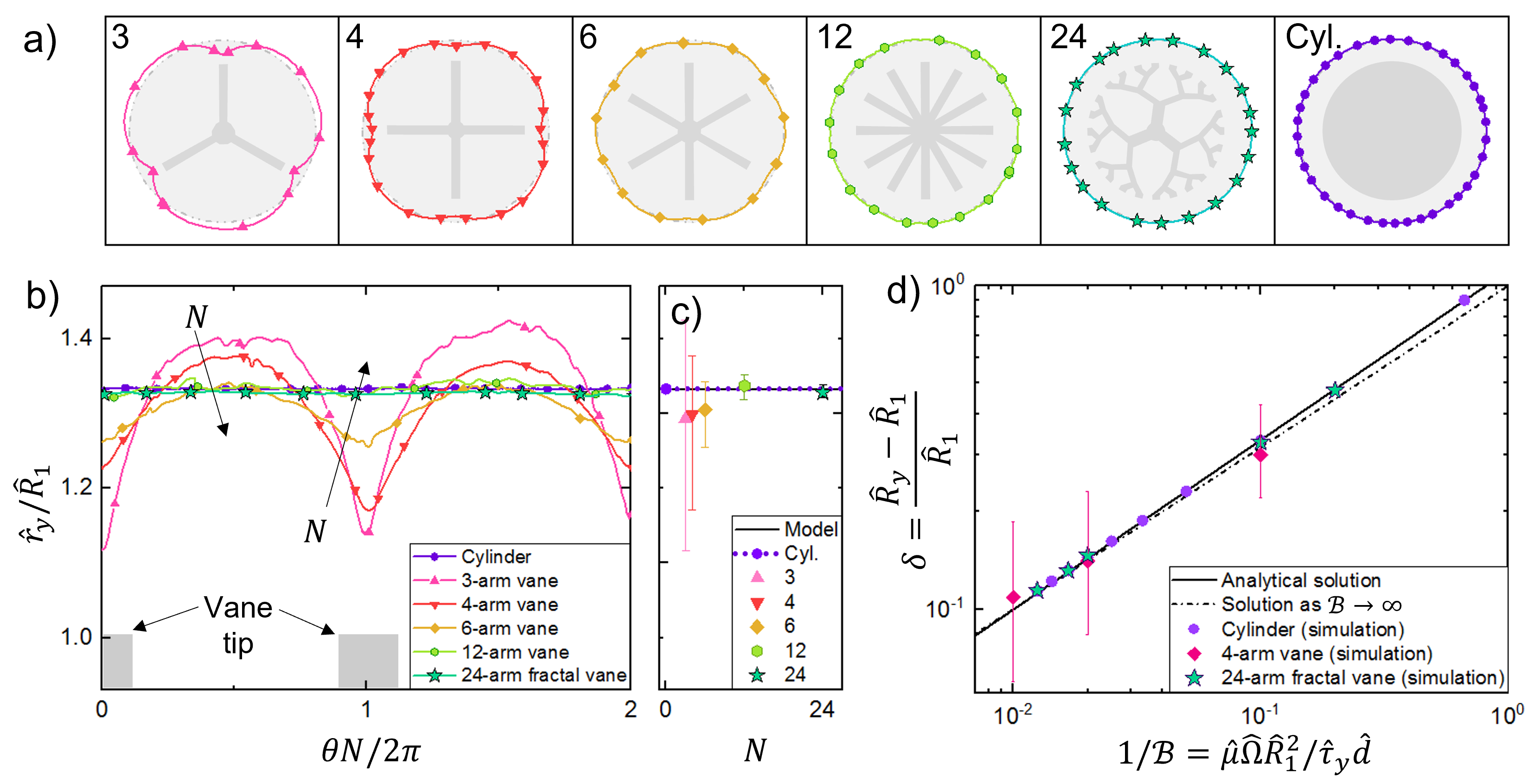}}%
    \caption{(a) Outer yield surfaces from the simulation are depicted for $\mathcal{B} = 10$ for a series of vanes with $N=3,4,6,12,24F$, and a cylinder, with a grey border showing the theoretical circular outer yield surface using Eq.~\ref{eq:VPBL}. (b) The solid vane structure impacts the outer yield surface, shown as radius normalized by the vane radius, $\hat{r}_y/\hat{R}_1$ as a function of the azimuthal angle between two arms. As the number of vane arms increases, the yield surface approaches that of a circle, (c) increasing in mean radius and decreasing in variability over the perimeter, where error bars indicate the full range of radius of the yield surface. (d) Location of the location of the outer yield surface with varying $\mathcal{B}$ for a cylinder and a 24-arm fractal vane agrees well compared to boundary layer theory for viscoplastic fluids in the limit of low $1/\mathcal{B}$ and at all points with an analytical solution as in Eq.~\ref{eq:PreciseBL} \cite{Landry2006}. All simulations are for Bingham fluids without slip. Due to high data density on yield surfaces, only 1/10 to 1/100 points from the simulation output values are shown as discrete points.}
    \label{fig:Outer_Yield_surface}
\end{figure}

The shape of the outer yield surface is compared for a series of vane structures in Fig.~\ref{fig:Outer_Yield_surface} for no-slip boundary conditions and $\mathcal{B}=10$. For vanes having $N=3,4$ and $6$ arms, the outer yield surface is quite sensitive to the structure of the vane arms even far from the vane tips, creating a lobed outer yield surface with $N$ protruding lobes (Fig.~\ref{fig:Outer_Yield_surface}(a,b)) between the vane arms. This general lobed structure has also been  documented previously in experiments \cite{Medina-Banuelos2019, Ovarlez2011}. When present, such non-axisymmetry of the flow introduces effects such as local extensional kinematic contributions that may  influence the rheological response of different fluid samples. These periodic lobes have notably reduced amplitude for $N\geq12$ as the profile becomes increasingly axisymmetric. While the mean radius of the yield surface, denoted $\hat{R}_y$ does not change markedly in the simulations of vanes with different $N$, the azimuthal variation in the radius reduces significantly with increasing $N$, as shown by the error bars in Fig.~\ref{fig:Outer_Yield_surface}(c).

As the Bingham number is increased the radial position of the outer yield surface moves progressively inwards towards the inner rotor $\hat{R}_y \to \hat{R}_1$, and the effects of shear are confined to a thin annular region of fluid. The position of the yield surface around a  cylindrical bob can be predicted analytically following Landry et al.~\cite{Landry2006} (see \ref{sec:Couette}). Their implicit solution for the (dimensionless) position of the yield surface $R_y(\mathcal{B})$ is rearranged and restated here: 
\begin{equation} \label{eq:PreciseBL}
\frac{2}{\mathcal{B}}=R_1 + 2~R_1~\log(R_y/R_1)-\frac{R_y^2}{R_1}.
\end{equation}
If we denote the dimensionless annular thickness  of residual sheared fluid by $\delta$: 
\begin{equation}\label{eq:VPBL}
\delta \equiv \frac{\hat{R}_y-\hat{R}_1}{\hat{R}_1},
\end{equation}
for the range of $\mathcal{B}$ considered here, this expression can be closely fit to a power law of the form $\delta\approx 1.11\mathcal{B}^{-0.52}$. Analysis of the evolution in the viscoplastic boundary layer thickness for a Bingham fluid at the limit of high $\mathcal{B}$ gives \cite{Boujlel2013,balmforth2017viscoplastic}, quite succinctly:
\begin{equation}
\delta \propto \mathcal{B}^{-1/2},
\end{equation}
which is in agreement with the power law fit. 

In Fig.~\ref{fig:Outer_Yield_surface}(d) these solutions are shown for a range of $\mathcal{B}$ in comparison to results from full 2D simulations of tools having cylindrical, $4$-arm, and $24$-arm fractal vane cross sections. The azimuthally-averaged mean radius of the outer yield surface $R_y = (2\pi)^{-1} \int_0^{2\pi} r_y(\theta)~d\theta$ is used when the outer yield surface is noncircular. Both the cylindrical bob and $24$-arm fractal vane agree almost perfectly ($< 0.5\%$ error) with the analytical prediction from Landry et al.~\cite{Landry2006} (Eq.~\ref{eq:PreciseBL}) for all values of $1/\mathcal{B}$ for which the yield surface is within the bounds of the cup, and also agree very well with viscoplastic boundary layer theory for small $1/\mathcal{B}$ (Eq.~\ref{eq:VPBL}). The $4$-arm vane has greater deviation from theory (up to $10\%$ for measured points), with substantially greater variation between the minimum and maximum yield surface points, as shown by the error bars in Fig.~\ref{fig:Outer_Yield_surface}(d). 

\begin{figure}[!h]
    \centering
    \makebox[\textwidth][c]{\includegraphics[width=0.7\textwidth]{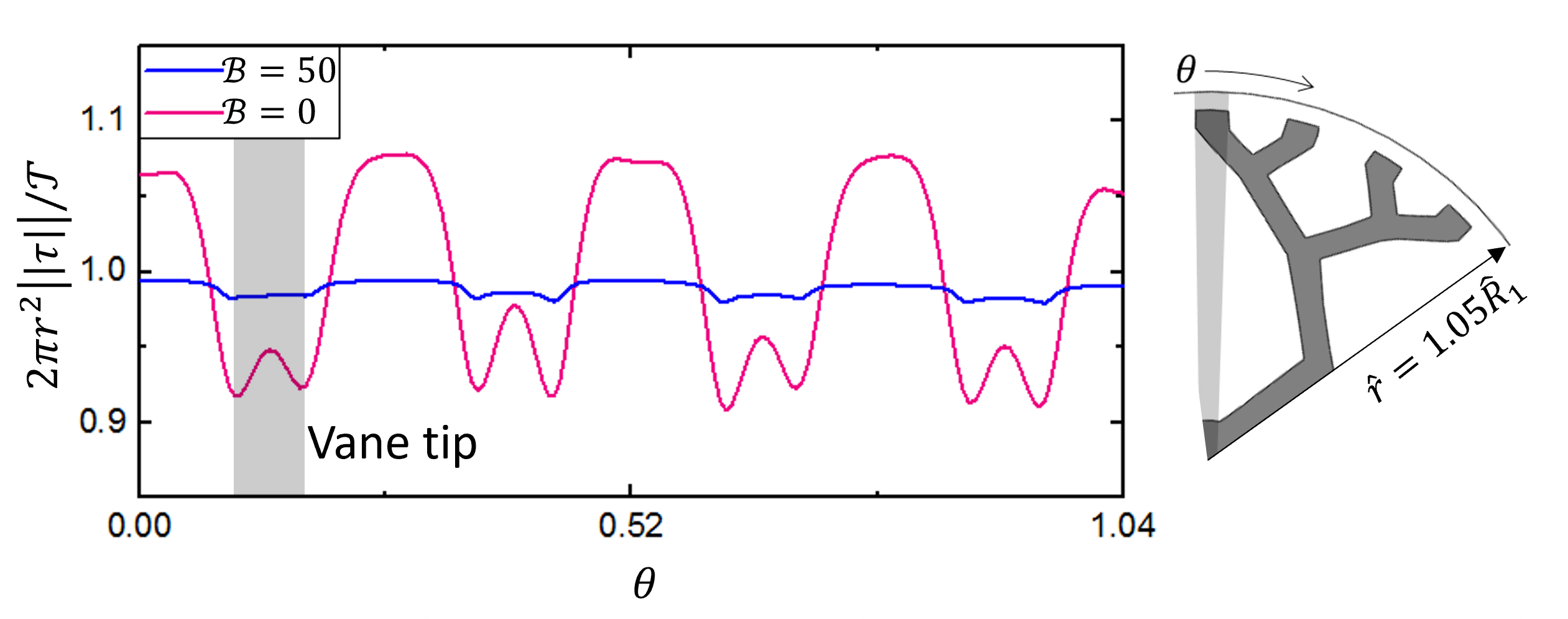}}%
    \caption{Azimuthal variation of the local stress field at a ring cut of $\hat{r}=1.05\hat{R}_1$. We compare the variations in the normalized stress contribution from simulations of a $24$-arm fractal vane for $\mathcal{B}= 0$ and $50$. The schematic indicates the precise orientation of the profile (with reference to the first vane tip), as the periodic variations in the stress are sensitive to the internal structure of the vane for $\mathcal{B}=0$.}
    \label{fig:stress-cutout-ring}
\end{figure}

The azimuthal variability of the yield surface indicates that along a circular arc at any constant radial coordinate the shear stress around any non-cylindrical vane will also vary. We compare the results of the local stress $||\tau||$ normalized by total torque $\mathcal{T}$ at a constant radius of $\hat{r}=1.05\hat{R}_1$ in Fig.~\ref{fig:stress-cutout-ring} for a $24$-arm fractal vane at $\mathcal{B}=0$ and $50$. A periodic fluctuating profile is apparent which is driven by the loci of the vane tips. The stress becomes increasingly sensitive to the vane structure at low-$\mathcal{B}$, having small variations in stress at each vane tip due to the changes in the internal spaces and local inclinations of the rigid no-slip boundaries within the structure. However, with increasing $\mathcal{B}$ these fluctuations markedly diminish from $18\%$ to $<2\%$ of the average stress, and the overall profile becomes nearly homogeneous. Similar results are obtained from analysis from each vane geometry, and serve to illustrate how even $4$-arm vane geometries can be accurate for viscometric measurement of yield-stress fluids. Greater errors will always be incurred when the vane is used to measure fluids without a yield stress ($\mathcal{B}=0$), but these will also be minimized by using a 24-arm fractal design rather than a 4- or 6-arm cruciform tool. 

\subsection{Total torque}
\begin{figure}[!h]
    \centering
    \makebox[\textwidth][c]{\includegraphics[width=1\textwidth]{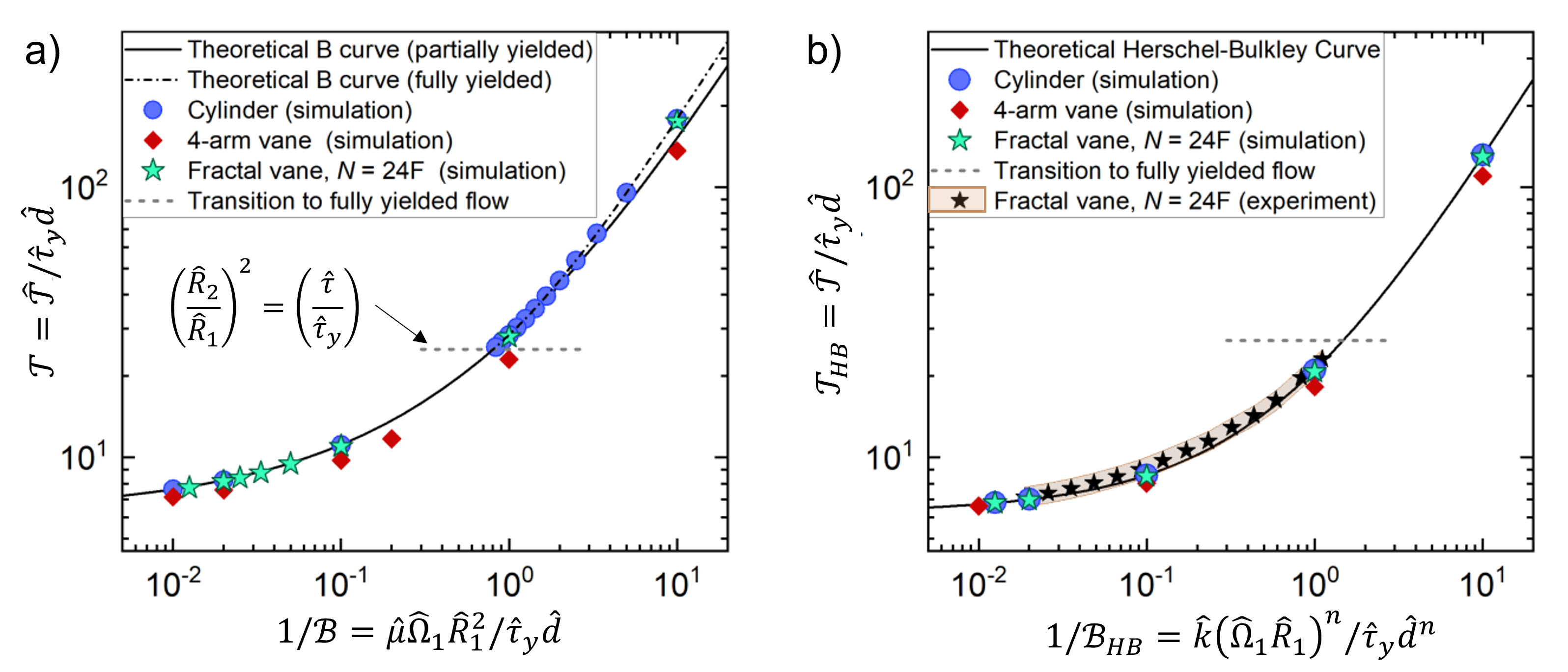}}%
    \caption{Dimensionless two dimensional torque $\mathcal{T}$ (as a torque per unit length of the vane) as a function of $1/\mathcal{B}$ for a large series of simulations of cylinders, 4-arm vanes, and 24-arm fractal vanes using (a) Bingham and (b) Herschel-Bulkley fluid models. Experimental data (shown by the black stars) was measured using a 24-arm fractal vane in Carbopol \cite{Owens2020} in a cup with $\hat{R}_2/\hat{R}_1 = 2$ and is presented with end effects removed using the approximation given in Eq~\ref{eq:endeffects}. A shaded error cloud indicates standard deviation of experimental measurement over three independent trials. The horizontal dashed line indicates the theoretical torque levels corresponding to the transition to fully yielded flow throughout the annular gap between the vane and the cup. The parameters for this Herschel-Bulkley fluid are yield stress $\hat{\tau}_y=109.5~Pa$, flow consistency $\hat{K} = 45~Pa.s^n$, and shear thinning index $n = 0.41$. }
    \label{fig:total_torque}
\end{figure}

Non-dimensional flow curves are shown as the two-dimensional torque $\mathcal{T}$ (as a torque per unit length of the vane) versus the dimensionless shear rate, $1/\mathcal{B}$ for simulations with the Bingham and Herschel-Bulkley constitutive models in Fig.~\ref{fig:total_torque}. The parameters for this Herschel-Bulkley fluid are yield stress $\hat{\tau_y}=109.5~Pa$, flow consistency index $\hat{K} = 45~Pa.s^{n}$, and shear thinning index $n = 0.41$. The experimental data included in Fig.~\ref{fig:total_torque}b uses a textured cup as depicted in Fig.~\ref{fig1} to mitigate slip effects at the outer wall. Analytical (dimensionless) flow curves are also shown as a reference for both fluid types. For the Herschel-Bulkley model we also show the sensitivity of the flow curve to the radial location of the outer yield surface. As we have noted above, because of the quadratic decay in the stress, when the average wall stress at the inner rotor surface is $\hat{\tau}_1 \le \hat{\tau}_y \big( \hat{R}_2/\hat{R}_1\big)^2$ then the fluid at the outer edge of the cup is unyielded and at rest (unless it slips). This limit is shown by the dashed line in Fig.~\ref{fig:total_torque}, and the analytical expression for the flow curve is derived in full in \ref{sec:HB} for this 'partially yielded' limit. 
At larger stress levels the fluid is yielded throughout the domain and this affects (weakly) the calculation of the effective shear rate distribution throughout the material, as can be seen by the broken line in Fig.~\ref{fig:total_torque}(a), which is adapted from a previous analysis \cite{Landry2006}. This analytical expression incorporating viscous dissipation throughout the gap agrees with our Bingham plastic simulations for both the cylindrical and 24-arm fractal vanes.
The results for the $4$-arm vane agree qualitatively with the rest of the data but fall systematically below and/or to the right of the reference curves. 

For the Herschel-Bulkley fluid, the theoretical curve is shown only for the case of partially yielded fluid because no general analytical solution could be found or derived despite much effort due to the exponent $n$ in the equations. However, the analytic solution remains in good agreement with the simulated torque values even above transition to fully yielded flow, reflecting the weaker contribution of viscous effects in this rate-thinning fluid when compared with a Bingham plastic fluid for which $n=1$. The close agreement between theory and fractal vane measurements at high shear stresses when the fluid is fully yielded across the gap ($\mathcal{B}^{-1} \gg 1$) remains to be verified experimentally. Further investigation may simplify the handling of raw data from such tools and potentially enable more facile adoption of vane/cup geometries with moderate gap ratios (here $\hat{R}_2/\hat{R}_1 = 2$ for all simulations) if a single equation can be used to handle all flow conditions without error beyond a specified threshold value. 

\section{Discussion}
\subsection{Experimental considerations} 
An experimental rheologist will note that the computational results presented in this study ignore some important practical considerations of vane rheometry including material sensitivity to (1) sample loading, and (2) interactions of a solid vane with aggregates or inhomogeneities within the sample itself. Both effects can contribute to error in rheometric measurement using any tool. It is for reasons of sample loading (1) that the relatively sparse $24$-arm fractal vane would be preferred over a vane with $24$ straight arms. Due to the ``cloaking" effects noted previously \cite{Chaparian2017}, the velocity and stress fields in the sheared fluid would be expected to be the same for a vane with $24$ outer contact points and any internal structure; however, the fractal design achieves this level of fluid/tool contact with a lower \textit{occluded area fraction} (OAF) resulting in penetration into samples with less material deformation.  Similarly, for a fluid sample that has large particles or aggregates, a textured cylinder would often be insufficient, due to wall slip or shear-induced segregation effects near the fluid/solid boundary. Although the flow kinematics are more idealized for many-arm vanes, penetration of solid aggregates into the finer-scale features may be restricted and fewer-arm vanes (e.g.~a 12-arm fractal vane) will facilitate sample loading and  enable more robust measurements. While vanes effectively prevent slip of material on the vane geometry itself, they cannot prevent shear banding, which is the formation of a temporary or permanent thin layer depleted of solids within the bulk material, and such inhomogeneities have been observed in imaging studies of concentrated emulsions and fibrous suspensions even while using vane geometries \cite{Ovarlez2011, Derakhshandeh2010}

\subsection{Constitutive relations and conversion equations}
\begin{figure}[!h]
    \centering
    \includegraphics[width=0.9\textwidth]{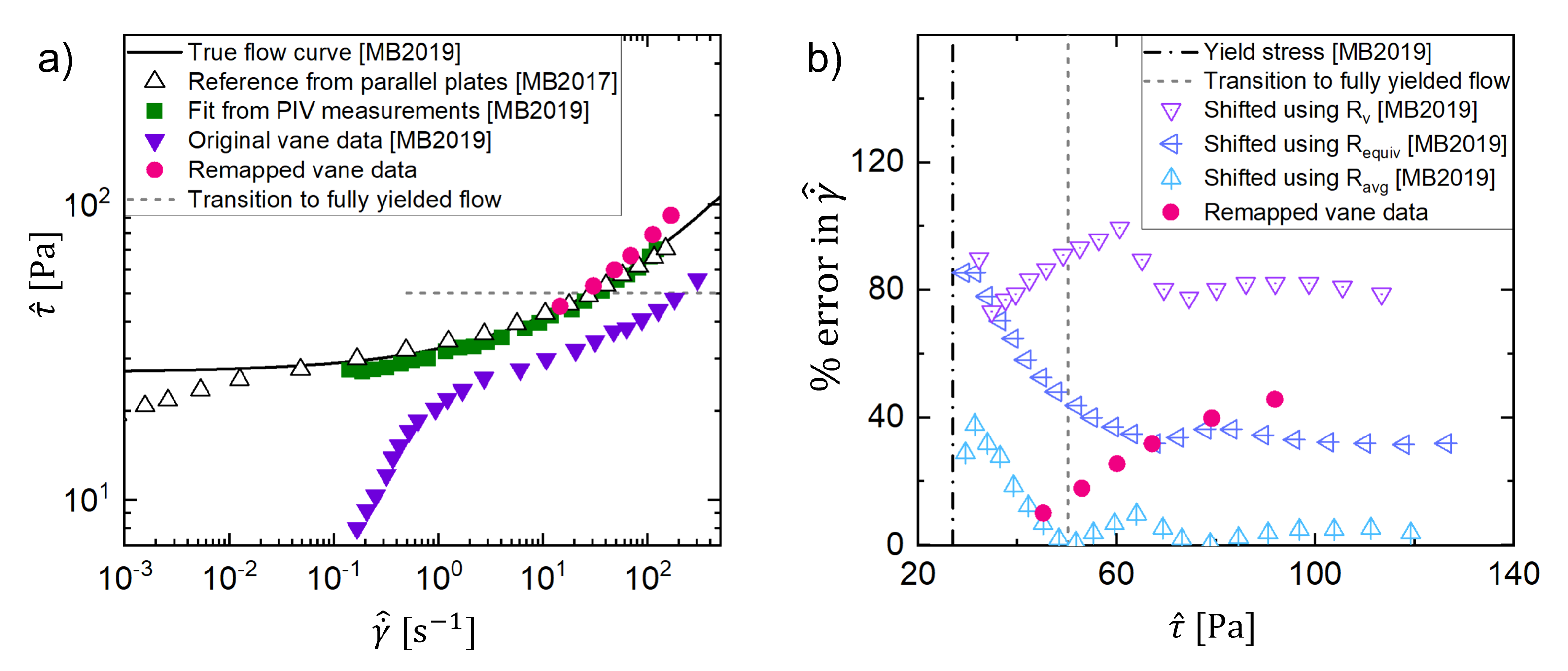}
    \caption{(a) Flow curves measured using different methods and equations on the same yield-stress fluid. (B) Error in data measured with the vane processed using different kinematic models including Eqs.~\eqref{eq:converttorque} and \eqref{eq:SR} and using different values for the effective vane radius. Specifically, $\hat{R}_1$ is the actual radius of the 6-arm vane, $\hat{R}_{equiv}$ is the effective radius calculated from measurements of a Newtonian fluid, and $\hat{R}_{avg}$ is the average of $\hat{R}_1$ and $\hat{R}_{equiv}$. Results taken directly from Medina-Ba{\~{n}}uelos 2019 \cite{Medina-Banuelos2019} without modification are noted with [MB2019] within the caption, and likewise [MB2017] for \citep{Medina-Banuelos2017}.}
    \label{fig15}
\end{figure}

%
The choice of tools used to measure material functions greatly influence the homogeneity of the flow kinematics generated in a rheometer. Meanwhile, the equations used to convert from the raw machine variables (torque, rotation rate) to true material properties (shear stress, shear rate) will have a substantial impact on the interpretation of all rheometric data collected with any tool. To illustrate this in vane rheometry specifically, we return to the data presented by Medina-Ba{\~{n}}uelos et al.~\citep{Medina-Banuelos2019} that was presented in Fig.~\ref{fig8}(a). This data set includes three independent measurements of the flow curve for the same test fluid measured with three different methods: (i) torsional rheometry using a parallel plate fixture with 150 grit sandpaper, which minimizes the effects of wall-slip and may thus be viewed as the `ground truth', to which the true flow curve is fit (as originally published in \cite{Medina-Banuelos2017} and replotted in \citep{Medina-Banuelos2019}); (ii) spatially-localized measurements across the gap (using the known radial variation of stress coupled with PIV to calculate the local shear rate), and finally (iii) conventional vane rheometry measurements. These different measurements of the material flow curve are shown in Fig.~\ref{fig15}(a).  

The PIV measurements (green squares) can produce a close approximation to the reference parallel-plate data. However this approach requires specialized optical equipment for flow imaging and transparent working fluids.  What we desire is an accurate process for converting the raw measurements of torque and rotation rate from the rheometer output to shear stress and shear rate at the vane perimeter. The original data reported in Medina-Ba{\~{n}}uelos et al.~\cite{Medina-Banuelos2019} with a standard 6-arm cruciform vane is thus shown here in two ways: (1)~after analysis with the originally-reported conversion equation (purple triangles), and (2)~after reprocessing the raw rotation rate and torque values (taken from Fig.~\ref{fig8}) with our equations~\eqref{eq:converttorque} and \eqref{eq:SR}. Note that the systematic deviation in the original vane rheometer data below a shear rate of approximately 1~s$^{-1}$ is reported to arise because of dramatic wall slip effects \cite{Medina-Banuelos2019}, and hence there is bound to be some uncertainty in our reprocessing of the original data due to the paucity of slip-free data available, coupled with the need for numerical computation of the derivative required for evaluating Eq.~\eqref{eq:SR}. 

The dashed boundary line in Fig.~\ref{fig15}(a) delineates the stress above which we would expect the fluid to become fully yielded throughout the gap for the specific cup geometry used \cite{Medina-Banuelos2019}. 
Below this stress level, the assumptions leading to Eq.~\eqref{eq:SR} are valid and the single conversion equation applies to mapping the measured rotation rate to the vane shear rate.
However, above this  point the fluid is fully yielded across the gap and boundary conditions change; therefore, a different equation for calculating the true shear rate at the rotating fixture would be appropriate. Multiple variants of this expression have been derived for fully yielded shearing flow between concentric cylinders. However the solutions depend on the constitutive model selected for the fluid \cite{Macosko1994_TCintro}, and incorporate various levels of complexity and accuracy depending on the assumptions made. For example, the thin gap assumption most readily simplifies calculations by removing the radial dependence of shear rate, but the resulting error already exceeds 3\% for a radius ratio $R=\hat{R}_1/\hat{R}_2=0.99$ and increases rapidly for larger gaps \cite{Macosko1994_TCintro}. For power law fluids, a Maclaurin expansion is typically used in deriving the relationship between shear rate and rotation rate, and different levels of error tolerance influence what order expansion is required, and thereby considerably affect the final form of the conversion equation \cite{Macosko1994_TCintro}. 
These approximate solutions are fully applicable to steady shearing flow in vane-and-cup geometries by invoking the \textit{Couette analogy} with an appropriately-selected effective radius, $\hat{R}_{\textit{eff}}$ \cite{Nguyen1987, Baravian2002, Estell2012}.

We explore the consequences of these decisions in more detail in Fig.~\ref{fig15}(b) (which are all calculated using the same input data for steady shear of the Carbopol fluid from \cite{Medina-Banuelos2019}). For data that fall within the range of partially-yielded flow (i.e.~to the left of the dashed boundary line), the level of error in the computed wall shear rate $\hat{\dot{\gamma}}$ varies widely. It can reach 100\% error (i.e.~a factor of 2) when using the vane radius $\hat{R}_1$ directly in the Couette analogy, and it remains above 80\% when using a standard calibration approach by measuring a Newtonian fluid to calculate the most commonly-used ``effective radius" $\hat{R}_{\textit{eff}} = \hat{R}_{equiv}$. The error may decrease below 40\%  using other post-measurement fitting methods (beyond standard calibration with Newtonian fluids); for example here $\hat{R}_{avg}$ is selected as the geometric average of $\hat{R}_1$ and $\hat{R}_{equiv}$. The error falls below 15\% error using our Eqs.~\eqref{eq:converttorque} and \eqref{eq:SR}; however, only a single  point in the original data set actually falls within the relevant range. At higher stresses, the outer yield surface in the fluid contacts the stationary cup surface, complicating accurate computation of the shear rate. The consequences of this error in determining the true shear rate are shown in the slight, but systematic, leftward shift of the remapped points (magenta circles) at high stresses in Fig.~\ref{fig15}(a). The same effect was also demonstrated in the numerical computations presented in  Fig.~\ref{fig:total_torque}(a).

This analysis, coupled with the detailed experimental investigation of Medina-Ba{\~{n}}uelos et al.~\citep{Medina-Banuelos2019}, indicate that slip-free measurements (for example using roughened or ribbed cups), as well as the careful use of accurate conversion equations (such as our Eqs.~\eqref{eq:converttorque} and \eqref{eq:SR}) -- which are applied within clearly-evaluated bounds of applicability -- are both essential constraints for generating accurate flow curves from vane rheometry. As might be anticipated, the viscometric flow generated between a cone and a plate provides the most accurate rheometric measurements (provided slip can be prevented). However this is often not possible for pastes, slurries, or other highly-filled or history-sensitive materials. In such cases vane fixtures are of broader utility and the computational rheological analysis presented here shows that vanes can still provide accurate rheological data provided slip is minimized and the raw data is analyzed in an appropriate way.  

\section{Conclusions}
Numerical simulations were used to study the full flow field of yield-stress fluids around vane tools with $N=3$ to $24$ arms over a wide range of Bingham numbers, $0 \le \mathcal{B} \le 100$. The results were compared with experimental measurements available in the literature in order to understand how the velocity and stress fields in the sheared fluid around the vane are affected by the design of the tool geometry.  The integrated moment of the computed stress field at the geometry boundary was also evaluated to compare with the output torque that would be measurable by a rheometer, as well as to highlight sources of error. 

We investigated the impact of the vane structure on the fluid velocity field, and found that few-arm vane designs ($N\leq6$) significantly perturb the flow away from  axisymmetric kinematics, whereas many-arm vanes ($N\geq12$) are successfully ``cloaked" in order to deform an almost axisymmetric annular ring of yield-stress fluid that is indistinguishable from the ideal deformation field arising around a rotating and slip-free cylindrical bob. At high Bingham numbers the evolution of this thin annular ring with increasing $\mathcal{B}$ is well-described by viscoplastic boundary layer theory.  

The end effects arising from the three-dimensional nature of a real vane tool were calculated in order to enable direct comparison between the 2D simulation results and experiments. We also investigated the influence of wall slip at the cup and vane surfaces on the final flow curve measured by a rheometer. Slip conditions on the convoluted vane surface did not impact measured overall torque $\mathcal{T}$, while slip conditions on the outer wall can have significant impact. Wall slip at the cup wall can be reduced or removed by texturing the outer wall, using methods such as sandblasting, affixing sandpaper of controlled grit size, or creating ribs by machining or 3D printing.

Finally, the azimuthal homogeneity of the shearing flow generated by a $4$-arm vane, a $24$-arm fractal vane and a cylindrical bob was quantified and compared. The $24$-arm vane and the cylinder influence the axisymmetry and  evolution of the outer yield surface in the same way and result in quantitatively equal values of the total torque.  However the $4$-arm vane generates much greater azimuthally-periodic perturbations in the shape of the outer yield surface, and noticeably smaller values of the torque for all values of Bingham number $\mathcal{B}$ simulated. When combined with experimental considerations such as ease of loading and low cost of fabricating precise structures using additive manufacturing, this study suggests that fractal vanes and textured or otherwise slip-free cup geometries can reduce slip artifacts and produce more accurate flow curve measurements for yield stress fluids.

\section*{Acknowledgments}
C.E.O. was supported by the United States Department of Defense (DoD) through the National Defense Science \& Engineering Graduate Fellowship (NDSEG) Program and an MIT MathWorks Engineering Fellowship. The authors also disclose a U.S. patent application assigned to MIT related to the fractal cross-section of fractal vanes \cite{FractalVanePatent}. Research on the rheometry of yield-stress fluids in the Non-Newtonian Fluid Dynamics Group at MIT is supported in part by a gift from the Procter \& Gamble Company. 

We thank Esteban Medina-Ba{\~{n}}uelos and Jos\'e P\'erez-Gonz\'alez for providing raw experimental torque data from their paper \cite{Medina-Banuelos2019}.  

\appendix
\section{Exact solution for circular Couette flow}\label{sec:Couette}

Using a cylindrical coordinate system with its origin at the centre of the circular vane ($\hat{R}_1$) and the cup ($\hat{R}_2$), we solve for a purely azimuthal flow, $\hat{\boldsymbol{u}}=\left( \hat{u}_r,\hat{u}_{\theta},\hat{u}_z \right) = \left( 0,\hat{u},0 \right)$. Therefore, the governing equation (\ref{govern}) reduces to,
\begin{equation}
\displaystyle\frac{\mbox{d}}{\mbox{d} \hat{r}} \left( \hat{r}^2 \hat{\tau}_{r\theta} \right)=0,
\end{equation}
which results in,
\begin{equation}\label{shear_stress_Couette}
\hat{\tau}_{r \theta}=A/\hat{r}^2,
\end{equation}
where $A$ is the constant of integration and will be calculated later with appropriate velocity boundary conditions. Hence, if the flow in the gap is only partially yielded then the position of the yield surface is $\hat{R}_y=\sqrt{\vert A \vert/\hat{\tau}_y}$.

The Bingham constitutive equation connects the shear stress and the azimuthal shear rate $\hat{\dot{\gamma}}_{r \theta}= \hat{r} \frac{\text{d}}{\text{d} \hat{r}} \left( \frac{\hat{u}}{\hat{r}} \right)$ in the yielded region as,

\begin{equation}
 \left(\hat{\mu}+\displaystyle\frac{\hat{\tau}_y}{\left\vert \hat{r} \frac{\text{d}}{\text{d} \hat{r}} \left( \frac{\hat{u}}{\hat{r}} \right) \right\vert} \right) \hat{r} \frac{\text{d}}{\text{d} \hat{r}} \left( \frac{\hat{u}}{\hat{r}} \right) = \frac{A}{\hat{r}^2},
\end{equation}
which can be rearranged considering that the velocity monotonically decays outwards from the bob towards the cup:
\begin{equation}
\frac{\text{d}}{\text{d} \hat{r}} \left( \frac{\hat{u}}{\hat{r}} \right) = \frac{1}{\hat{\mu}} \left( \frac{\hat{\tau}_y}{\hat{r}}+\frac{A}{\hat{r}^3}\right).
\label{shear rate definition}
\end{equation}
Hence, the velocity profile in the yielded region is,
\begin{equation}
\hat{u} = \frac{1}{\hat{\mu}} \left( \hat{\tau}_y \hat{r} \log(\hat{r}) - \frac{A}{2 \hat{r}} + C \hat{r} \right),
\end{equation}
and the constants of integration $A$ and $C$ can be found from boundary conditions,
\begin{equation}
\hat{u} \left(\hat{R}_1\right) = \hat{R}_1 \hat{\Omega}_1~~\&~~u (\hat{R}_o) = 0,
\end{equation}
where $\hat{R}_o$ is the locus of the zero velocity contour; this can be either the edge of the yield surface $\hat{R}_o = \hat{R}_y$ (in the case that the whole gap is not yielded) or the outer cup wall, $\hat{R}_o = \hat{R}_2$ (when the bob rotates fast enough to yield all of the fluid across the entire gap width). Hence, $\hat{R}_o = \min{\left(\hat{R}_y , \hat{R}_2 \right)}$. In the latter case, some analytical progress is feasible which gives,
\begin{equation}
A = \frac{2\hat{R}_1^2 \hat{R}_2^2}{\hat{R}_2^2 - \hat{R}_1^2} \left[ \hat{\tau}_y \log \left( \frac{\hat{R}_1}{\hat{R}_2} \right) - \hat{\mu} \hat{\Omega}_1 \right].
\end{equation}
Therefore, the shear stress at the bob can be calculated from Eq.~\eqref{shear_stress_Couette} and also can be converted to calculate the local shear rate across the gap by substituting into the Bingham constitutive model.

\section{Theoretical Dimensionless Flow Curves for Fig.~\ref{fig:total_torque}}\label{sec:HB}

The solid analytical lines on Fig.~\ref{fig:total_torque} displaying torque as a function of rotation rate in the case of partially yielded flow are analytical solutions calculated from the Herschel-Bulkley model in Fig.~\ref{fig:total_torque}(a), with the special case of $n=1$ for the Bingham model solution in Fig.~\ref{fig:total_torque}(b). Details of the derivation are presented here for interested readers. The Herschel-Bulkley model is 
\begin{equation}
\hat{\tau} = \hat{\tau}_y + \hat{K}\hat{\dot{\gamma}}^n
\label{eq:Appendix-HB-equation}
\end{equation}
with specific constants for our Carbopol of $\hat{\tau}_y=109.5$~Pa, $\hat{K}=45$~Pa.s$^n$, and $n=0.41$. The caret, $<\hat{\cdot}>$, refers to dimensional quantities. For any given stress above the yield stress, the shear rate from Eq.~\eqref{eq:Appendix-HB-equation} is then
\begin{equation}\label{eq:A1.1}
\hat{\dot{\gamma}} = \left(\frac{\hat{\tau}-\hat{\tau}_y}{\hat{K}}\right)^{1/n}
\end{equation}
for $\hat{\tau}\geq\hat{\tau}_y$. In partially yielded flow, the measured rotation rate of the vane fixture can be used to evaluate the shear rate at the edge of the vane fixture by Eq.~\eqref{eq:SR} \cite{Nguyen1987}. This expression is repeated here in an adjusted form by noting that at $\hat{r}=\hat{R}_1$,
$d\log \hat{\mathcal{T}}/d\log \hat{\Omega} 
= d\log \hat{\tau}/d\log \hat{\Omega} 
= (\hat{\Omega}~d\hat{\tau})/(\hat{\tau}~d\hat{\Omega})$: 
\begin{equation}\label{eq:A1.2}
\hat{\dot{\gamma}} = \frac{2\hat{\tau}}{d\hat{\tau}/d\hat{\Omega}}.
\end{equation}
Note that derivation of Eq.~\eqref{eq:A1.2} includes the assumption that the shear rate (and the fluid velocity) reaches zero at some radius within the gap in order to evaluate the change in integration variable using the Leibniz rule \cite{Nguyen1987}.  Combining Eqs.~\eqref{eq:A1.1} and \eqref{eq:A1.2}, rearranging and integrating we obtain
\begin{equation}
\label{eq:integrate-stresses}
\int_{\hat{\Omega}(\hat{\tau}\rightarrow\hat{\tau}_y)}^{\hat{\Omega}(\hat{\tau})}{d\hat{\Omega}} 
= \hat{\Omega}(\hat{\tau}) - \hat{\Omega}(\hat{\tau}_y) 
= \frac{1}{2}\int_{\hat{\tau}\rightarrow\hat{\tau}_y}^{\hat{\tau}}{
\frac{1}{\hat{\tau}'}
\left(\frac{\hat{\tau}'-\hat{\tau}_y}{\hat{K}}\right)^{1/n}
d\hat{\tau}'}
\end{equation}
noting that as $\hat{\tau}\rightarrow\hat{\tau}_y$, then $\hat{\Omega}(\hat{\tau})\rightarrow0$ in slip-free flow. We solve for the rotation rate in terms of the stress: 
\begin{equation}
%
\hat{\Omega}(\hat{\tau}) = \frac{1}{2\hat{K}^{1/n}}\left[n(\hat{\tau})^{1/n}F_1\left(\frac{-1}{n},\frac{-1}{n};\frac{n-1}{n};\frac{\hat{\tau}_y}{\hat{\tau}}\right) - \pi\hat{\tau}_y^{1/n}\csc\left(\frac{\pi}{n}\right)\right]
%
%
\end{equation}
where $F_1$ is the Gaussian hypergeometric function and this expression is valid for $\hat{\tau}>\hat{\tau}_y$; for $\hat{\tau}\leq\hat{\tau}_y$, $\hat{\Omega}=0$. If $n=1$ and $\hat{K}=\hat{\mu}$, this analysis simplifies to describe a fluid with the Bingham plastic model and the solution to Eq.~\eqref{eq:integrate-stresses} is
\begin{equation}
\hat{\Omega} = \frac{1}{2\hat{\mu}}
\left(\hat{\tau}-\hat{\tau}_y
\left[1-\log\left(\frac{\hat{\tau}}{\hat{\tau}_y}\right)\right]\right). 
\end{equation}
This is again valid for $\hat{\tau}>\hat{\tau}_y$; for $\hat{\tau}\leq\hat{\tau}_y$, $\hat{\Omega}=0$. For plotting these solutions onto Fig.~\ref{fig:total_torque}, the shear stress $\hat{\tau}$ is transformed back to give the torque $\hat{\mathcal{T}}$ using Eq.~\eqref{eq:converttorque}. 


\bibliography{bibtex}

\end{document}